\documentclass[journal,12pt,onecolumn]{IEEEtran}
\usepackage{lmodern,times,amsmath,amssymb,epsfig,epstopdf,color,subfigure,booktabs,multirow,turnstile,graphicx,multicol,textcomp,graphicx,colortbl}
\usepackage[utf8]{inputenc}
\usepackage[acronym]{glossaries}
\usepackage{url}
\usepackage{txfonts}
\usepackage{siunitx}
\usepackage{subfiles}
\usepackage{todonotes}
\usepackage{mathrsfs,dsfont}
\usepackage{amssymb,mathrsfs}
\usepackage[mathscr]{euscript}
\usepackage{epstopdf} 
\usepackage{float}
\usepackage[noadjust]{cite}
\usepackage[scaled=.9]{helvet}
\usepackage{grffile}
\usepackage{wrapfig}
\usepackage{amsmath}
\usepackage[final]{changes}
\usepackage{flushend}
\graphicspath{{./}}
\DeclareMathOperator*{\argmin}{argmin}

\newacronym{MPC}{MPC}{Multi-path component}
\newacronym{Tx}{Tx}{transmitter}
\newacronym{IoT}{IoT}{Internet of Things}
\newacronym{Rx}{Rx}{receiver}
\newacronym{PL}{PL}{Path Loss}
\newacronym{PLE}{PLE}{Path Loss Exponent}
\newacronym{MaMIMO}{MaMIMO}{Massive-Multiple-In-Multiple-Out}
\newacronym{MIMO}{MIMO}{Multiple-In-Multiple-Out}
\newacronym{A2G}{A2G}{Air-to-ground}
\newacronym{A2A}{A2A}{Air-to-air}
\newacronym{LOS}{LOS}{Line-of-Sight}
\newacronym{NLOS}{NLOS}{Non-Line-of-Sight}
\newacronym{UAV}{UAV}{Unmanned Aerial Vehicle}
\newacronym{LS}{LS-fading}{Large-Scale fading}
\newacronym{SS}{SS-fading}{Small-Scale fading}
\newacronym{mmWave}{mmWave}{millimeter wave}
\newacronym{AUE}{AUE}{Aerial User Equipment}
\newacronym{UE}{UE}{User Equipment}
\newacronym{ABS}{ABS}{Aerial Base Station}
\newacronym{BS}{BS}{Base Station}
\newacronym{SINR}{SINR}{Signal-to-Interference-and-Noise-Ratio}
\newacronym{SNR}{SNR}{Signal-to-Noise-Ratio}

\begin{document}
%
\title{Tutorial on UAVs:  A Blue Sky View on Wireless Communication}
%
%
%

\author{Evgenii Vinogradov*, Hazem Sallouha, Sibren De~Bast, Mohammad Mahdi Azari, Sofie Pollin
\thanks{KU Leuven, Department of Electrical Engineering - ESAT, Leuven Belgium.}

\thanks{*Corresponding Author E-mail: evgenii.vinogradov@kulueven.be }
}

%
%

\markboth{Tutorial on UAVs:  A Blue Sky View on Wireless Communication}%
{Vinogradov E. \MakeLowercase{\textit{et al.}}}
%



\maketitle

\begin{abstract}
The growing use of Unmanned Aerial Vehicles~(UAVs) for various applications requires ubiquitous and reliable connectivity for safe control and data exchange between these devices and ground terminals.
Depending on the application, UAV-mounted wireless equipment can either be an aerial user equipment (AUE) that co-exists with the terrestrial users, or it can be a part of wireless infrastructure providing a range of services to the ground users. 
For instance, AUE can be used for real-time search and rescue and/or video streaming (surveillance, broadcasting) and Aerial Base Station (ABS) can enhance coverage, capacity and energy efficiency of wireless networks.
In both cases, UAV-based solutions are scalable, mobile, easy and fast to deploy.
However, several technical challenges have to be addressed before such solutions will become widely used. 
In this work, we present a tutorial on wireless communication with UAVs\added{,} taking into account a wide range of potential applications. 
The main goal of this work is to provide a complete overview of the main scenarios (AUE \replaced{and}{or} ABS), channel and performance models, compare them, and discuss open research points. 
This work is intended to serve as a tutorial for wireless communication with UAVs, which gives a comprehensive overview of the research done until now and depicts a comprehensive picture to foster new ideas and solutions while avoiding duplication of past work.
We start by discussing the open challenges of wireless communication with UAVs.
To give answers to the posed questions, we focus on the UAV communication basics, mainly providing the necessary channel modeling background and giving guidelines on how various channel models should be used. 
Next, theoretical, simulation- and measurement-based approaches\added{,} to address the key challenges for AUE usage\added{,} are presented. 
Moreover, in this work, we aim to provide a comprehensive overview on how UAV-mounted equipment can be used as a part of a communication network. 
Based on the theoretical analysis, we show how various network parameters (for example coverage area \deleted{of ABSs}, power efficiency, or user localization error \added{of ABSs}) can be optimized. 
\end{abstract}

\begin{IEEEkeywords}
UAV, drone, A2G, ABS, AUE, aerial, channel modeling, shadowing
\end{IEEEkeywords}

%
\IEEEpeerreviewmaketitle
\section{Introduction}

\acrfull{UAV}-enabled solutions, systems, and networks are considered for various applications ranging from military and security operations to entertainment and telecommunications \cite{6852096, namuduri2017, Valavanis2014, 5600072, safedroneware, 6564694, mozaffari2016efficient, 8038869, 7470933, 7317490,6599064}.  
\acrshort{UAV}s (or drones) are becoming more and more popular owing to their flexibility and potential cost efficiency in comparison with conventional aircrafts.
Business Insider Intelligence (UK) published results of their market research \cite{BI} where they predict that sales of UAVs \added{will} surpass \$12 billion per year by 2021, which is up by a compound annual growth rate (CAGR) of 7.6\% from \$8.5~billion in 2016. 
Commercial Drone\deleted{s} shipments will reach 805,000 in 2021, a CAGR of 51\%. 
\par
The global \acrshort{UAV} (including military drones) payload market value is expected to reach \$3 billion by 2027 (the payload consists of all equipment carried by UAVs such as cameras, sensors, radars, communications equipment, and others). Radar\deleted{s} and communication\deleted{s} equipment dominate the global UAV payload market with a market share of close to 80\%, followed by cameras and sensors segment with around 11\% share \cite{SDInt}.
\par
Since \deleted{the} drones become more functional, reliable, and affordable, UAV-based solutions for new markets start being competitive.
In \cite{BI}, the value of drones, \added{sorted} by industry\added{,} in 2021 \replaced{is}{was} estimated as: Infrastructure (\$45.2~B); Security (\$10~B); Media and Entertainment (\$8.8~B); Telecommunications (\$6.3~B). 
\par
The market size and dynamics resulted in a significant interest from the academia and industry in \acrshort{UAV}-based solutions.
In this work, we give a comprehensive overview of the progress and challenges that drone-enabled wireless communications face nowadays.
\subsection{Aerial Wireless Communication}
\begin{figure}
\centering
  \includegraphics[width=.5\linewidth]{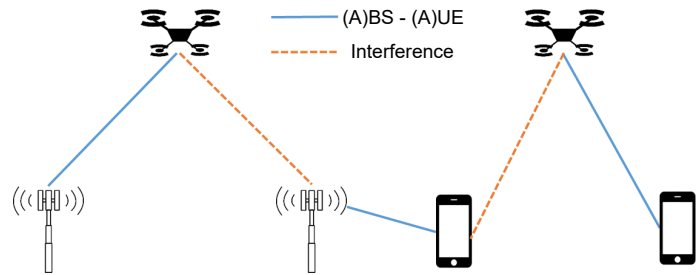}
  \caption{Aerial User Equipment and Aerial Base Station scenarios}
  \label{fig:AUE_ABS}
\end{figure}
Drones can act as \deleted{a} flying \acrfull{UE} (see Figure~\ref{fig:AUE_ABS}, left side). 
For instance, \replaced{a}{an} \acrshort{UAV} equipped with a camera (and other necessary sensors) can provide a cost efficient solution for surveillance, inspection, and delivery.
In this case, so-called \acrfull{AUE} has to co-exist with ground users and exploit existing infrastructure (such as cellular networks) to transfer collected information to the operator on the ground with certain reliability, throughput, and delay\added{,} depending on the application requirements.
\par
On the other hand, the use of \added{a} UAV-mounted \acrfull{BS} is an alternative future technology that can provide power-efficient wireless connectivity for ground users (see Figure~\ref{fig:AUE_ABS}, right side).
Due to its mobility and flexibility, an \acrfull{ABS} can dynamically provide additional capacity on-demand.
This solution can be used by service providers both for dynamic network densification, fast network deployment in an emergency situation, or temporary coverage of an area.
Moreover, due to the favorable propagation conditions the localization service precision can be significantly improved.
Note that the interference to the ground infrastructure also has to be taken into account.

\begin{figure}
\centering
  \includegraphics[width=.7\linewidth]{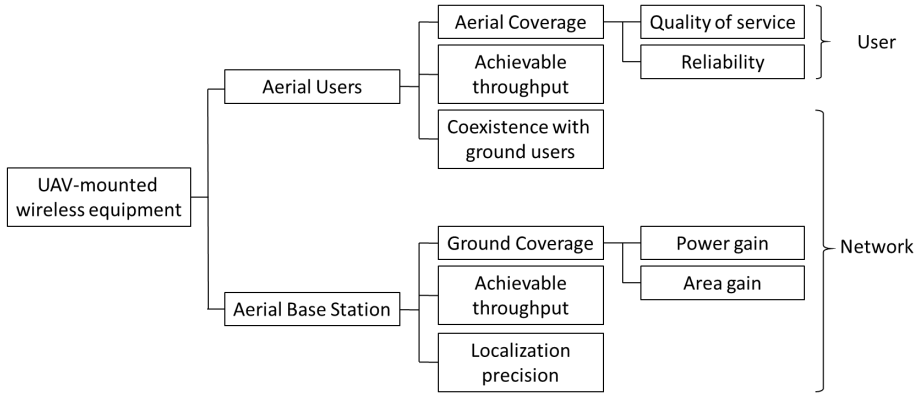}
  \caption{Important performance metrics for different UAV-enabled wireless communication systems}
  \label{fig:UAv_overiew}
\end{figure}
UAV-enabled wireless communication networks have to be studied taking into account appropriate channel models\footnote{Note that a complete channel model consists of \acrfull{PL}, \acrfull{LS}, and \acrfull{SS} models considering the 3D location of both terminals, environment (rural, sub-urban, urban, etc.), frequency, and other physical link parameters.}, antenna configurations, (A)UE\deleted{s} and (A)BS\deleted{s} densities\added{,} and other network parameters. 
\added{A} coverage and reliability analysis is very different in both scenarios (see Figure~\ref{fig:UAv_overiew}). 
\par
The first scenario requires a good link from the AUE to at least one of the \acrshort{BS}\added{s} deployed\added{,} typically at the rooftop level.
At the same time, the other \acrshort{BS}s become interferers and cause a performance drop.
In this scenario, the main focus is at the \acrshort{AUE} performance, however, its coexistence with the terrestrial \acrshort{UE}s and the network infrastructure has to be studied.
\par
The second scenario requires a good link between \replaced{all}{any} \added{of the } ground \acrshort{UE}\added{s} \replaced{to one of}{from} the multiple \acrshort{ABS}s.
The radio wave propagation conditions are completely different for aerial channels when compared with the terrestrial channels.
Moreover, due to the aforementioned difference in \added{the} ground nodes (BS and UE) height for the \acrshort{AUE} and \acrshort{ABS} scenarios, the propagation conditions significantly change.
Another important issue is that the aggregate interference to and from the air is different. 
Consequently, the techniques and services designed for classical wireless networks (e.g. localization) have to be adapted to the new 3D paradigm.
It is obvious that completely different approaches are necessary for \acrshort{AUE} and \acrshort{ABS} performance analysis.
\begin{table}
\centering
    \caption{UAV-related works}\label{tab:overview}

    \begin{tabular}{c|c|c|c|c|c}
    \toprule

	     &\multicolumn{2}{c|}{PL: With} &{PL: Without}&SS&Measurement\\ 
   	     &\multicolumn{2}{c|}{LOS/NLOS} &{LOS/NLOS }&-fading&-based\\ 
         &\multicolumn{2}{c|}{separation} &{separation}&&\\
        & Fixed& 3D  & Fixed & \\
        &PLE& variant &PLE&  \\
        && PLE &&  \\
                \midrule
 \multirow{5}{*}{Channel modeling}&\cite{6162389,6566747 }&\cite{3gpp_uav, 7037248}&\cite{7835273, 6187152, 7407385, 7357682}&\cite{7842372, 3gpp_uav}&\cite{7842372, 3gpp_uav}\\
 &&\cite{7842372,6863654}&\cite{7501562, 7870687}&\cite{7835273, 6187152, 7407385, 7357682}&\cite{7835273, 6187152, 7407385, 7357682}\\
 &&\cite{3gpp-summary, 4917538}&&\cite{3gpp-summary, 6492100}&\cite{3gpp-summary, 6492100}\\
   &&&&\cite{7501562,7870687}&\cite{7501562,7870687}\\
   &&&&\cite{4917538, 8287893}&\cite{4917538, 8287893}\\
		\midrule
                 \midrule
        \multirow{3}{*}{AUE}&\cite{azari2017coexistence}&\cite{azari2018reshaping, azari2018cellular}&\cite{8301389, liu2018comp}&\cite{azari2017coexistence, azari2018cellular}&\cite{6852096,Asadpour:2013:LDD:2535372.2535409}\\
        &&\cite{8337920}&&\cite{8337920,azari2018reshaping}&\cite{lte_in_the_sky, Bergh2015AnalysisOH}\\
        &&&&&\cite{QC_report,8301389}\\
        \midrule
        \multirow{3}{*}{ABS}&&\cite{azari2017ultra,Azari2017CoverageMF}&\cite{azari2016optimal,7470932}&\cite{azari2016optimal,azari2017ultra  }&\multirow{4}{*}{\cite{7470932}}\\
        &&\cite{azari2016joint, hayajneh2016}&\cite{lte_in_the_sky,wu2018joint}&\cite{azari2016joint,Azari2017CoverageMF }\\
        &&\cite{7417609,mozaffari2015}&\cite{zhan2018energy,Zeng}&\cite{hayajneh2016, lte_in_the_sky}\\
        &&\cite{8038869}&&\\
\bottomrule      
		\end{tabular}
\end{table}
\subsection{Literature Review and Important Research Questions}

UAV communication performance analysis has attracted a large body of research so far. 
Papers focus on the channel modeling only, or on system performance evaluation both for AUE\added{s} or ABS\added{s}.
A detailed overview of the published research \deleted{works} is given in Table~\ref{tab:overview}.
We can broadly classify the approaches: the theoretical analysis (e.g. using the published channel models as a tool for AUE or ABS performance estimation) versus measurement based research.  
\par
The channel model should accurately reflect the environment seen by the wireless link to ensure a correct and accurate performance analysis of the AUE and ABS communication. 
As it can be seen, the main trend is to develop and use \deleted{the} channel models differentiating two kinds of propagation: \acrfull{LOS} and \acrfull{NLOS}. 
Moreover, the most elaborated of these models use an altitude-dependent \acrfull{PLE}.
Several AUE and ABS performance analysis papers consider \acrshort{SS}\added{,} which indeed makes those works more complete and realistic. 
Here we highlight main research challenges, summarize the previously published literature\added{,} and discuss its limitations.
\par
\subsubsection{Channel modeling}\label{par:channel} 
Channel modeling is one of the fundamental issues for designing any wireless technology.
Ground-to-ground networks are understood and modeled very well.
In contrast, aerial communication channels have been investigated much less.
The main challenge in \acrfull{A2G} channel modeling is the complexity of 3D environment\added{s} and a large set of parameters that must be considered: \acrshort{PL}, \acrshort{LS}, and \acrshort{SS} behavior depends on the environment type (urban, rural, etc.), \acrfull{Tx} and \acrfull{Rx} heights, incident and/or elevation angles, the carrier frequency, \acrshort{LOS} probability, etc.
Future techniques like \acrfull{mmWave} and \acrfull{MaMIMO}  still need to be studied in depth to improve \replaced{their}{the} channel models in \replaced{the}{their} aerial context.
\par
\added{The} aeronautical channel models survey in \cite{6187152} discusses many channel modeling efforts, but these models can \added{only} be used for aircrafts flying higher than the operating altitude of a typical commercial drone.
Several models of an aerial channel for lower heights have been proposed in literature (see Table~\ref{tab:overview}).
Well-known log-distance and two-ray \acrshort{PL} models were parameterized via measurement campaigns. Some works (e.g. in \cite{7842372}) in fact, parameterized two separate models: for LOS and NLOS cases, whereas \cite{7835273, 6187152, 7407385, 7357682} did not draw this distinction. 
In \cite{6863654}, an analytical model was presented. 
Measurement campaigns suggest that \acrshort{SS} is usually Rayleigh or Nakagami distributed.
\replaced{Most of the time, published work uses a statistical approach, whereas \cite{miao2018a} proposes to use a Ray-tracing tool to model the channel.}{Mostly, the published works are using the statistical approach, whereas \cite{miao2018a} proposed to use  Ray-tracing tool for modeling the channel.}
\par
In this tutorial, we provide a comprehensive overview of existing channel models that can be directly used for practical UAV applications as well as for research.
We draw \added{an} explicit separation between different propagation slices (or echelons): the ground level (below 10~m and 22.5~m for suburban and urban environments, respectively), obstructed \acrshort{A2G} channel (10 - 40~m and 22.5 - 100~m), and high-altitude \acrshort{A2G} channel (40 - 300~m and 100 - 300~m).
Then we provide the model parameters for each slice.
Moreover, we show examples of the practical implementation of these models for simulation based performance estimation of several drone-enabled scenarios.

\subsubsection{Aerial user equipment} 
AUE performance \replaced{investigation}{estimation} is needed since \replaced{the majority of}{the popular} wireless technologies are not designed to operate in \acrshort{A2G} links\replaced{.}{,} \replaced{Therefore, the}{so their} performance is unpredictable in such \replaced{peculiar}{particular} environments.
\deleted{Moreover, the current cellular infrastructure are optimized for ground users.  As a result, the performance for aerial users using cellular networks has to be studied.} 
\replaced{Recent reports have concluded}{It was shown in many papers} that \replaced{interference is the main source of \acrshort{AUE} performance degradation}{the main issue limiting \acrshort{AUE} performance is interference.} 
Consequently, the main challenge is to estimate the performance taking into account a complex propagation channel (see above) and the terrestrial network topology (BS density and location, Tx power, 3D antenna patterns). \added{Furthermore, }coexistence of AUE with ground \acrshort{UE} must be investigated \deleted{as well}. 
\par
Numerous research activities in this topic are well summarized in the surveys \cite{7317490, 7463007, 6564694, Kim2018, 8337920,lte_in_the_sky} where the requirements for quality-of-service, data, connectivity, adaptability, security etc. were quantified.
In \cite{7317490}, possible aerial network architectures were inspected, whereas \cite{7463007} reports the characteristics and requirements for UAV-networks for several promising civil applications. 
\deleted{No analytical frameworks were provided.}
Some separate use-cases were investigated in \cite{6564694, Kim2018}: it was underlined that for entertainment and virtual reality applications, the communication channel can become the main limiting factor.
AUE performance was analyzed in \cite{8337920} \deleted{via}  using a simulator consisting of \acrshort{PL} channel model and 3D antenna patterns.
Based on measurements \added{introduced} in \cite{lte_in_the_sky}, it was concluded that the interference is one of the main problems for AUE scenarios. 
The measurement-based performance estimation (see Table~\ref{tab:overview}) considers specific environments and does not provide a unified framework for the AUE performance analysis.
The \replaced{results presented}{work} in \cite{azari2017coexistence,8301389, liu2018comp} use a channel model with a fixed \acrshort{PLE} (independent of altitude), \replaced{however}{moreover}, only \cite{azari2017coexistence} considers the effect of \acrshort{SS}.
\par
While existing literature considers many important issues, it has some limitations.
The surveys are limited to isolated UAV application use-cases, so that the information is fragmented.
A work \replaced{that considers}{considering} all possible aspects and approaches to the \acrshort{AUE} wireless communication performance estimation (from theoretical analysis to experimental results) is missing, \added{to the best of our knowledge}.
\deleted{In addition to the performance study for existing cellular networks in combination with aerial users, the main limitations on the performance for aerial users have to be investigated.}
This knowledge is \added{the} key to optimize future cellular networks \replaced{while considering}{for} aerial \replaced{users{nodes}}.
\par
In this tutorial we give an overview of the theoretical state-of-the-art.
We proceed with the simulation based performance estimation of a cellular network serving an \acrshort{AUE}. 
Next, we complete the study by giving an overview of relevant measurement campaigns detailing the currently achieved UAV communication for existing communication technologies such as LTE and Wi-Fi.
We show that by considering the impacts of the altitude, environments, antenna configuration, and network density, the \acrshort{UAV} position potentially can be optimized in order to achieve the highest coverage possibility.

\subsection{Aerial communication infrastructure} 
\acrshort{ABS}s, aerial relays \replaced{on}{in} ad~hoc networks, and \deleted{aerial anchors} are \added{a} promising addition to the conventional wireless networks. 
It is vital to understand whether \replaced{the use of}{using} ABSs \replaced{is}{will be} beneficial. 
Consequently, the main challenge is to propose guidelines for the optimal 3D positioning, ABS deployment density, and path-planning (when the UAV carrying the communication equipment is mobile) maximizing the advantages of ABS over the conventional terrestrial BSs. The complete analysis should take into account the complex propagation channel, antenna patterns, practical limitations (e.g. power consumption, weight or maximum flight altitude) or required network performance metrics.
\par
Multiple papers listed in Table~\ref{tab:overview} consider mostly specific use cases and/or communication aspects without providing a theoretical framework that could be used for a complete analysis of an aerial communication network.
Survey \cite{7470932} investigates the optimal positioning of a helikite-mounted ABS as well as considers more practical aspects such as dimensions and required power. 
The survey provides measurement results, but the analytical results are limited: a very basic channel model is used and no framework is proposed.
An overview of ABS-aided networks is given in \cite{7470933}, however, it also lacks \added{the} theoretical analysis and appropriate attention to the used channel models.
\par
In this work, we aim to provide a comprehensive overview on how UAV-mounted \added{user} equipment can be used as a part of the communication network. 
Based on the theoretical analysis, we show how \replaced{one}{we} can optimize the coverage area of \acrshort{ABS}s. 
The size of this area is a function of several parameters, e.g. UAV height, transmit power and antenna tilt. 
The presented framework allows \deleted{to} \replaced{estimating}{estimate} the effect of these parameters on the \acrshort{ABS} system and optimal deployment configuration.
Moreover, the influence of multiple \acrshort{ABS}s can be investigated.
\replaced{Similar to}{Just like with} terrestrial BSs, multiple ABSs interfere with each other. 
Therefore, the density of these ABSs must be studied to maximize the coverage of the network, while minimizing the interference.
Additionally, the parameters important for ABS's deployment \replaced{is}{can be} studied to optimize the localization service they will provide. 
For example, we analyze how the height of the UAV influences the positioning error of the localization system.
\subsection{Tutorial Organization} 
Summarizing, the main contributions of this article is the overview and tutorial on the wireless communications for \acrshort{UAV}.
Our purpose is to provide an introduction of the UAV wireless communications to the reader not working in this topic but interested in getting a general introduction to the subject.
This tutorial also targets researchers wishing to get a comprehensive background before working on the subject.
The main goal is to gather and systematize the vast, but fragmented state-of-the-art research contributions regarding most important aspects of \acrshort{UAV}-enabled wireless communications networks.
\par
The tutorial is structured as follows:
\begin{itemize}
\item Section~\ref{Sec:2} provides an overview of the motivating \acrshort{UAV} applications for both scenarios: UAV as the UE or UAV as integral part of the communication system.
\item Section~\ref{Sec:Channel} is focused on the UAV communication basics, mainly giving an overview of the main \acrfull{A2A} and \acrshort{A2G} channel models taking into account \acrshort{PL}, \acrshort{LS}, and \acrshort{SS}. These models consider the 3D location of both terminals, environment, frequency, and other important physical link parameters. The guidelines on how the models can be applied are given.
\item Section~\ref{Sec:UE} discusses the communication performance for \acrshort{UAV}s as an \acrshort{AUE} scenarios, focusing on analytical, simulation, and measurement results.
\item Section~\ref{Sec:System} discusses two promising \acrshort{UAV} applications when they are a part of the network infrastructure. First, we present an analytical communication performance estimation for \acrshort{ABS}s followed by performance simulations.
Next, we discuss the localization performance, for scenarios where the UAV is a part of a localization system.
\item Section~\ref{Sec:future} provides an overview of the future research directions and challenges.
\item Section~\ref{Sec:conc} concludes the paper.
\end{itemize}
\section{Wireless Communication for UAVs: Use Cases}\label{Sec:2}
A wide range of use cases can be imagined for \acrshort{UAV}s. 
When analyzing the performance of UAV communication, it is important to divide the use cases in two classes of scenarios, one where the UAV is used as a mobile terminal and the second one where the UAV is a part of the wireless communication infrastructure.
In the first case, the UAV is considered as an \acrshort{AUE}, where the data generated by the payload of the UAV needs to be transmitted to serve an application. 
In the second case, the UAV is exploited as a part of the communication system, for example, as a mobile base station or anchor for localization of \acrfull{IoT} devices. 
Below, we first describe some scenarios where the UAV is a mobile wireless terminal, and then we focus on scenarios where the UAV is part of the infrastructure.
\subsection{
Unmanned Aerial Vehicles as User Equipment
}
  \begin{figure}[!t]
  \centering
       \includegraphics  [width=.5\linewidth] {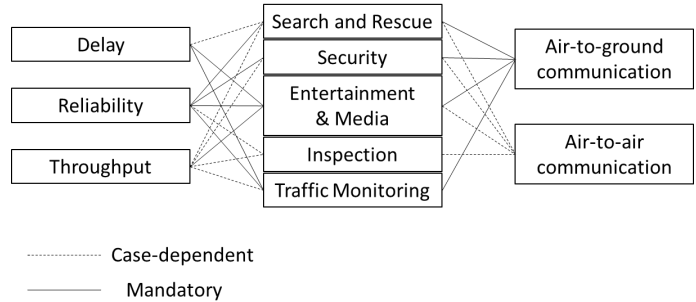} 
      \caption{Qualitative requirements for various use-cases when UAV acts as an UE}
      \label{fig:AUE_req}
  \end{figure}
Naturally, drones can act as users of the wireless infrastructure.  
To properly use drones as \acrshort{AUE}, the requirements for the communication between drones and ground infrastructure should be adapted to the considered application (see Figure~\ref{fig:AUE_req}). 
For instance, it is obvious that for search and rescue, the reliability is vital. 
In the case when the UAV-mounted camera is used to help a ground-based operational center, the delay and throughput requirements are high.
However, due to the possibility of augmenting the UAV with autonomous on-board functionality such as positioning and person detection, the drone can also only transmit the coordinates of the person. 

Many studies are dedicated to the feasibility of \acrshort{AUE} usage in existing cellular networks (see Table~\ref{tab:overview}). 
It turns out that the interference becomes the main limiting factor for the wide deployment of the cellular-connected drones due to the fact that current cellular networks were designed for ground users whose operations, mobility, and traffic characteristics are substantially different from the \acrshort{AUE}. 

There are key differences between drone-mounted and ground \acrshort{UE}s.
The propagation conditions for \acrshort{A2G} and terrestrial channels are completely different due to nearly \acrshort{LOS} communications between ground \acrshort{BS}s and \acrshort{AUE}s. 
On one hand, it improves the received signal level, but in return the interference from the other \acrshort{BS}s also grows. 
Moreover, the ground users are in general less dynamic than UAVs, therefore, the techniques used currently in the cellular networks might require significant modification to enable the highly mobile users.

In the case of an \acrshort{AUE}, it may need to transfer the information about to the security operation center.
This information might vary from telemetry-like signals to a more demanding video stream.
Let us describe the most promising \acrshort{AUE} applications.

\subsubsection{Search and Rescue}
Drones have been used for searching victims \cite{5600072}. 
Due to the freedom of mobility, they can be used in various dangerous scenarios. 
The authors of \cite{avalangeRescue} proposed a specialized UAV platform to search and rescue people in case of avalanches, which is capable of locating survivors fast, and hence ensuring a maximal survival chance. Furthermore, when fitted with the right sensors, UAVs can also be used for indoor and outdoor urban rescue missions \cite{urbanRescue}. As a results, from  May  2017  through  April  2018,  DJI  has  counted 65 people  who  were  rescued  from  peril  by  use  of  a  drone \cite{DJI}.
Recently, drones also proved to be a perfect tool to prevent casualties. The Los Angeles Times reported \cite{LATimes}, that infrared drone footage, taken from high altitude, informed a group of firefighters near Yosemite National Park, that they were facing seven spot fires instead of just one they had been aware of. Moreover, after a fire burned, drones are ideal for damage assessment as it has been done after wild fires in Greece \cite{Greece}.

\subsubsection{Security}

UAVs are able to optimize their path quickly and complete complex missions due to their high mobility, which makes them attractive for various security and safety-related applications.
Equipped with the right sensors and actuators, UAVs can monitor an area for illegal activities via video surveillance.
For surveillance purposes, \cite{Perera2018} describes a method to detect humans on aerial footage and estimate the pose and trajectory of the subjects.
In addition, multiple drones can be controlled by one ground control station to dynamically secure a large area \cite{Perez2013}.
Other UAVs can be used in order to detect \cite{VinoRadar} and/or intercept  \cite{aeroguard} malicious drones, e.g. drones equipped with a net can catch malicious UAVs. 



\subsubsection{Entertainment and Media}
The entertainment business was one of the first businesses to widely use drones for the production of TV shows and movies. But for these applications, the video is stored on board on the \acrshort{UAV}. In the future, there are plans to use drones for live broadcasting and AR/VR applications \cite{Kim2018}, this requires a reliable high-throughput link to safely transmit the stream from the drone to the ground operating center. 

Another option is to use drones as broadcasting relays. For big sports events, like bikes races, motorcycles are used to film the athletes. The stream gets relayed via a helicopter to the control station of the TV station. Operating this helicopter is not cost efficient, therefore an \acrshort{UAV} could be equipped with relaying infrastructure \cite{freitas2010} and autonomously optimize its position to retransmit the video streams.
\\
Another example of an entertainment application for drones comes from Intel, which uses drones to make animated light shows \cite{ShootingStar}. These Intel \textit{Shooting Star} drones are fitted with LED bulbs and fly in formation to form breath taking displays. In order to coordinate the movement from more than a thousand drones, a reliable and low-latency communication system is needed that sends the correct commands to all the drones. 
\\
The next application for entertainment purposes came from Dolce \& Gabbana. During their Winter 2018 showcase in Milan, they demonstrated their handbags and purses by flying them down the catwalk using drones \cite{Dolce}. 

\subsubsection{Inspection}

Autonomous \acrshort{UAV} can easily take over when highly repetitive flights (such as a periodic industrial infrastructure inspection) have to be performed. An example is the inspection of wind turbine blades for damage and wear. When checking the blades of an entire wind turbine park, the pilot can loose the concentration. This leads to dangerous situations where the pilot could crash or miss a damaged spot on the turbines blades. Autonomous drones rely on cameras and software to do these routine inspections and do not have the problem of fatigue, therefore the inspections will become more secure and effective. The same procedure can also be applied on other big industrial installations e.g. petrochemical plants and cooling towers \cite{microdrones}. Several projects and companies are already dedicated to this use case, like SkySpecs \cite{skyspecs} and SafeDroneWare \cite{safedroneware,safedroneware_dancing}.


\subsubsection{Traffic Monitoring}
Owing to their dynamic and multidisciplinary characteristics, drones have become increasingly popular for traffic monitoring. This interest resulted in several research activities. One of the first related surveys \cite{6564694} concludes that UAVs are proven to be a viable and less time-consuming alternative to real-time traffic monitoring and management, providing the eye-in-the-sky solution to the problem.
\par
The main functions that an aerial-based traffic monitoring system should have are i) object (pedestrian, car etc.) detection \cite{7163593,rs9040312}, ii) tracking \cite{WANG2016294} and iii) analysis of the collected information (e.g. the flow density and dynamics) \cite{rs10030458}.
In addition, UAVs can be deployed on demand in a fast and dynamic way. This can be extremely interesting for sudden traffic jams due to accidents or large events. The monitoring services at these locations will help to avoid traffic jams at these locations in the future.

\subsection{Unmanned Aerial Infrastructure }
  \begin{figure}[!ht]
  \centering
       \includegraphics  [width=.7\linewidth] {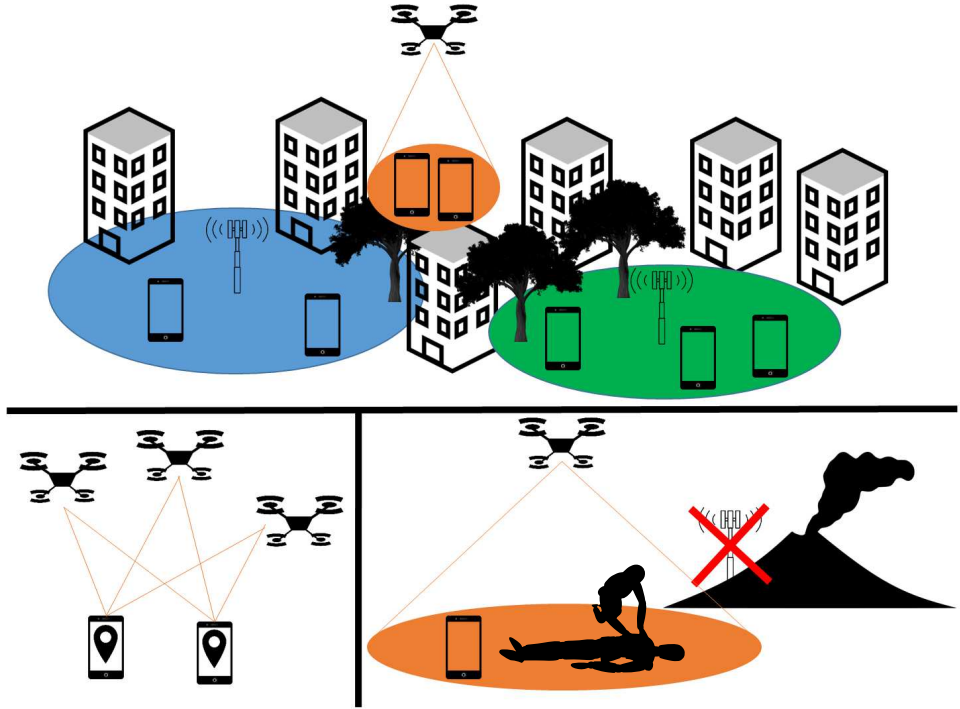} 
      \caption{Aerial Base Station use cases. Top: Future communication networks; Bottom, Left: Localization service; Bottom, Right: disaster scenario}
      \label{fig:disaster}
  \end{figure}
\subsubsection{Aerial Base Stations: Future Telecommunication Networks}

The demand for high-speed wireless access has been incessantly growing last years.
Smart-phone traffic will exceed PC traffic by 2021: traffic from wireless and mobile devices will account for more than 63 percent of total IP traffic by 2021 \cite{Cisco}.
The interest in enhancing the capacity and coverage of existing wireless cellular networks has led to the emergence of new wireless technologies, which include ultra-dense small cell networks, \acrshort{mmWave} communications and \acrshort{MaMIMO}. They often are collectively referred as the next-generation 5G cellular systems. 
\par
We believe that UAV-mounted base stations will become an important component of the 5G environment due to the ability of providing on-demand connectivity to the users at little additional cost \cite{mozaffari2015} as shown in Figure \ref{fig:disaster}~(top). 
Meanwhile, a \acrshort{mmWave} \acrshort{ABS} mounted on a UAV \cite{xiao2016} can naturally establish \acrshort{LOS} connections (which is vital for the wireless links at these frequencies) to ground users. 
In its turn, combining \acrshort{mmWave} communications and \acrshort{MaMIMO} can be an attractive solution to provide high capacity wireless transmissions.

\subsubsection{Aerial Base Stations: Public Safety During Natural Disasters}

  Natural disasters such as earthquakes, fires, and floods often yield to severe damage or complete destruction of the existing terrestrial communication networks. In particular, cellular base stations and ground communication infrastructure is often damaged and overloaded during natural disasters (see Figure \ref{fig:disaster}, bottom right). It is obvious that there is a need for public safety communication solutions for search and rescue operations. The potential broadband wireless technologies for public safety scenarios include 4G Long Term Evolution (LTE), Wi-Fi, satellite communications and dedicated public safety systems \cite{6599064}. 
  \par
  An \acrshort{ABS} can be seen as a part of a robust, fast, and capable emergency communication system enabling effective communications during public safety operations\cite{hayajneh2016}. For instance, AT\&T deployed an LTE cell site on a helicopter to connect residents after a disaster in Puerto Rico in 2017 \cite{ATT_COW}. However helicopters have a high operating cost. While UAVs can easily fly and dynamically change their positions to ensure full coverage to a given area within a minimum possible time and at a low operating cost. Therefore, the UAV-mounted base stations can be seen as the key enabler for providing fast and ubiquitous connectivity in public safety scenarios.

\subsubsection{Aerial Anchors: Positioning for Ground Users}
Another advantage of using drones as part of the wireless infrastructure, is that they can be used for localization of terrestrial nodes (see Figure \ref{fig:disaster}, bottom left). Aerial positioning systems have the advantage of having a higher \acrshort{LOS}-probability than a terrestrial \acrshort{BS}, which increases the possible positioning-accuracy. \cite{aerial_anchors} envisions a localization system based on aerial anchors, where terrestrial users can localize themselves using the Received Signal Strength (RSS) of three or more static UAVs. \cite{Pinotti2016} proves that these drones can also be replaced by one single drone if the users are static. Other proposed systems make use of time-based approach. Both of them have advantages and disadvantages. RSS based systems can use already available data to calculate the position of the device hence, this strategy is very power efficient. This comes with the drawback of being less accurate. Therefore, this option is ideal for static battery powered \acrshort{IoT} devices, since their main limitation is battery-stored energy. Moreover, since these devices are not mobile, the accuracy error can be lowered by averaging over time. These characteristics make this system a viable option over  Global Positioning System (GPS) to position static IoT devices. Time-based systems can be used to complement GPS signals to further enhance the positioning accuracy. This is possible due to their higher accuracy than the RSS-based systems. However, they do need a very accurate clock signal at the transmitter and the receiver, which requires a substantial amount of energy to operate. Both technologies can have potential to enhance the current available positioning and to achieve higher energy efficiency or higher accuracy.
\section{Fundamentals of UAV Communication}
\label{Sec:Channel}
In wireless communication networks, the propagation channel is the medium between the transmitter and the receiver. 
It is obvious that its properties influence the performance of wireless networks. Next generation \acrshort{UAV}-enabled networks design (i.e., algorithms development) is impossible without the knowledge of the wireless channel. 
Consequently, radio channel characterization and modeling in such innovative architectures becomes crucial to evaluate the achievable network performance. 
\par

\begin{figure}
\centering
     \includegraphics  [width=.5\linewidth] {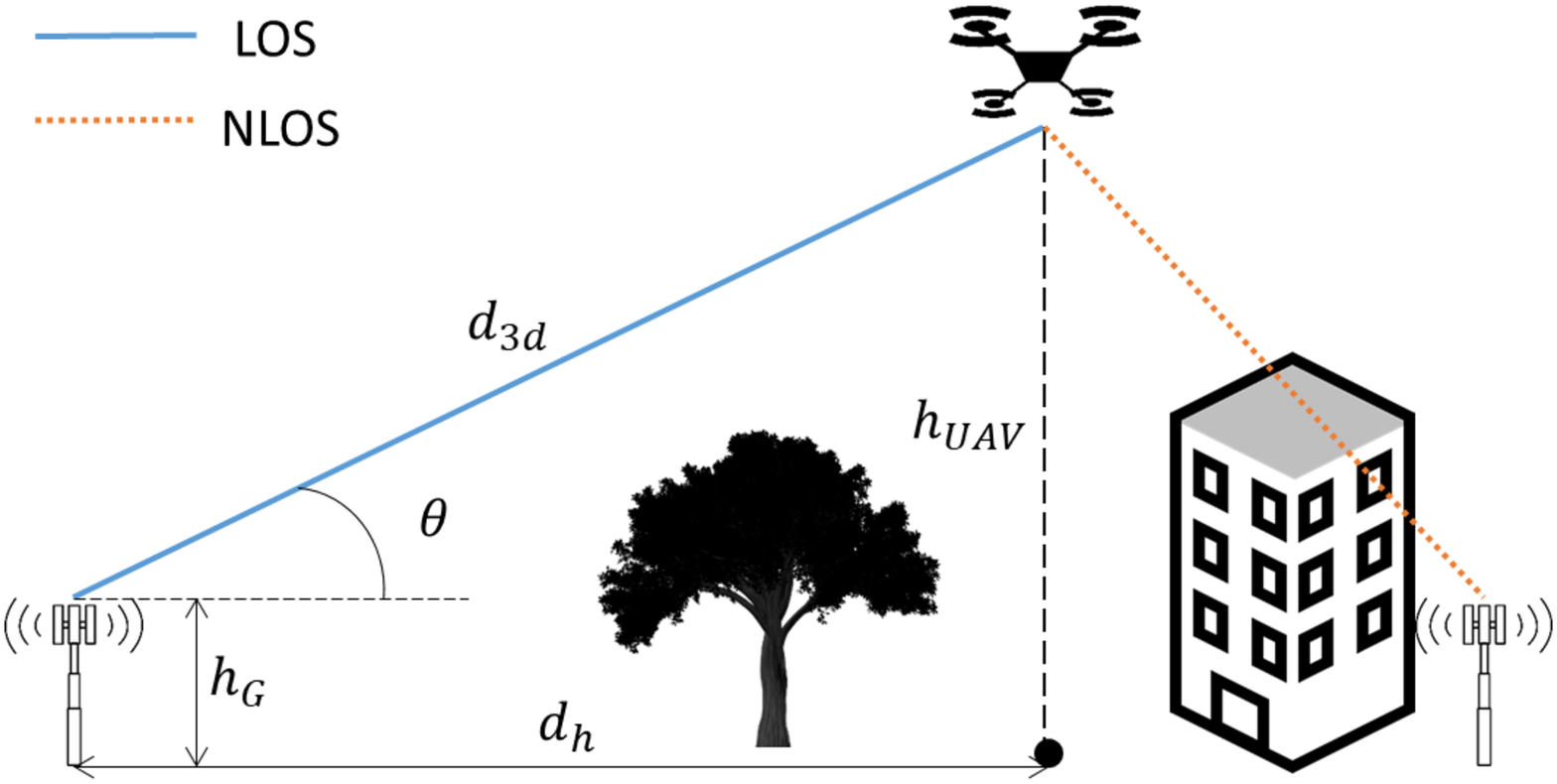} 
    \caption{Air-to-Ground propagation}
    \label{fig:A2G_param}
\end{figure}

The vast majority of the channel modeling efforts is dedicated to terrestrial radio channels.
Unfortunately, wireless communication with \acrshort{UAV}s cannot rely on these models.
The nature of \acrshort{A2G} channels implies a higher probability of \acrshort{LOS} propagation.
This results in a higher link reliability and lower transmission power.
Even for \acrshort{NLOS} links, power variations are less severe than in the terrestrial communication networks due to the fact that only the ground-based side of the link is surrounded by the objects that affect the propagation.
Figure~\ref{fig:A2G_param} illustrates \acrshort{A2G} propagation channel and introduces the main geometrical parameters as well as drawing the important distinction between LOS and NLOS channels\footnote{There is no trivial dependency between any of these parameters and the performance metrics described in the following sections. Any change of the link geometry results in a complex change of the network operation.}.
On the other hand, the \acrshort{UAV} mobility causes high rates of change.
Modeling of these changes is challenging due to arbitrary mobility patterns and complex operational environment.
Doppler shift caused by the UAV motions has to be taken into account as well. 
\par
Apart from the channel itself, there are other factors that influence the received power strength such as: airframe shadowing, self-interference from the on-board devices and antenna characteristics. 
\par
An accurate channel model of \acrshort{A2G} channels is vital for design and optimal deployment of the communication networks including \acrshort{UAV}s as its nodes.
In this section we will discuss a number of research questions and recent efforts dedicated to this important issue.

\subsection{Background}
\begin{figure}
\centering
     \includegraphics  [width=.5\linewidth] {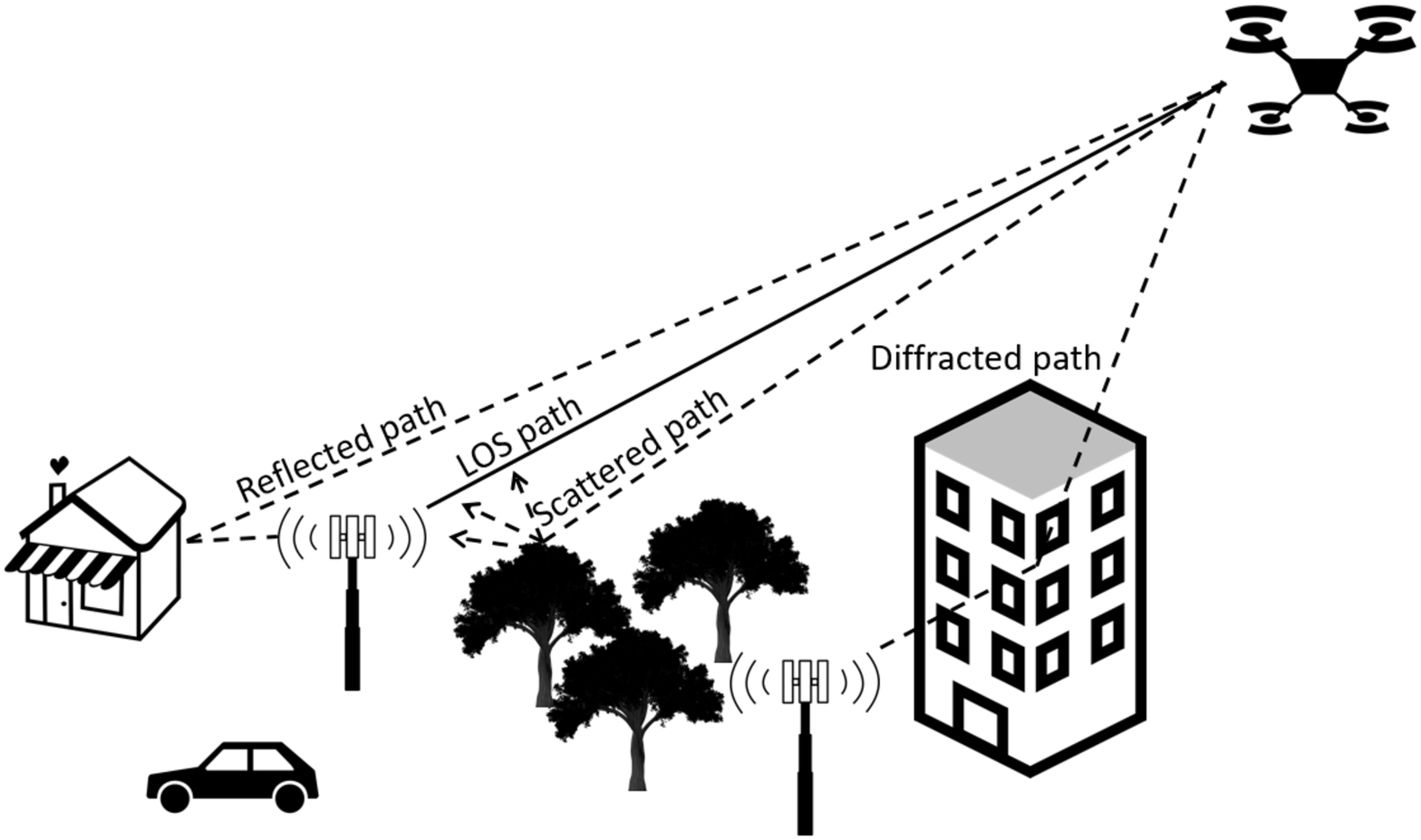} 
    \caption{Air-to-ground propagation}
    \label{fig:A2G_channel}
\end{figure}
The transmitter radiates electromagnetic waves in several directions. Waves interact with the surrounding environment through various propagation phenomena before they reach the receiver. 
As illustrated in Figure~\ref{fig:A2G_channel}, different phenomena such as specular reflections, diffraction, scattering, penetration or any combination of these can be involved in propagation \cite{Oestges-Clerckx}. 
Therefore, multiple realizations of the transmitted signal, often termed as \acrfull{MPC} arrive to the Rx with different amplitudes, delays and directions.
The resulting signal is the linear coherent superposition of all copies of the transmitted signal, which can be constructive or destructive depending upon their respective random phases.
\par
\begin{figure}
\centering
     \includegraphics  [width=.5\linewidth] {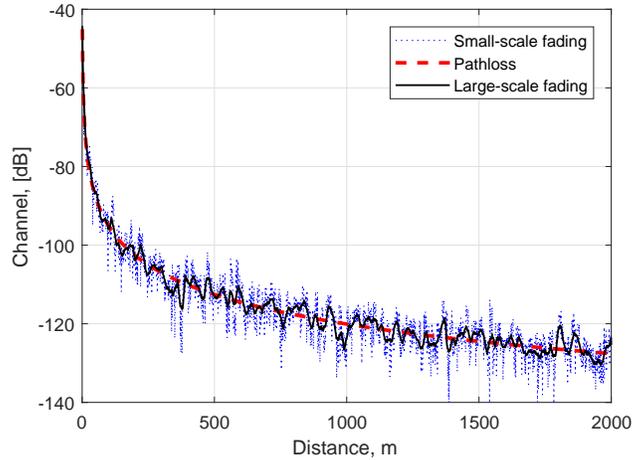} 
    \caption{Channel components}
    \label{fig:channel}
\end{figure}

Typically, radio channels can be represented as a superposition of several separate fading mechanisms:
\begin{equation}\label{eq:channel}
H=\Lambda+X_{LS}+X_{SS},
\end{equation}
where, $\Lambda$ is the distance dependent \acrshort{PL}, $X_{LS}$ is the \acrshort{LS} (also known as shadowing) consisting of large scale power variations caused by the environment, and $X_{SS}$ is the \acrshort{SS} (see Figure~\ref{fig:channel}). 
\par
Depending on the altitude, different channel models (or their parameters such as \acrshort{PLE} or LOS probability) must be used due to the obvious difference in experienced propagation conditions. The airspace is often separated into three propagation slices (or echelons): 
\begin{itemize}
\item Ground level: below 10~m or 22.5~m for suburban and urban environments, respectively \cite{3gpp_uav}. These channels can be modeled by using terrestrial channel models since the aerial node altitude is below the rooftop level. Mostly the NLOS propagation is expected.
\item Obstructed \acrshort{A2G} channel: 10 - 40~m and 22.5 - 100~m for suburban and urban environments, respectively. These  channels experience a higher LOS probability than the ground channels, however, it is not 100\%. Consequently, a large variation of the received power around the mean \acrfull{SNR} levels is observed.
\item High-altitude \acrshort{A2G} channel: 40 - 300~m and 100 - 300~m. Above certain altitude (depending on the environment), all channels are in \acrshort{LOS}, so the propagation is close to the free-space case. Consequently, only LOS channel model can be used. Moreover, no \acrshort{LS} is expected.
\end{itemize}
Additionally,  \acrshort{A2A} channels are mostly experiencing LOS propagation and similar to the high-altitude \acrshort{A2G} channels. Note that in this case, the UAVs mobility can be significantly higher which causes larger Doppler shifts.
\par
Next, let us describe models of the components presented in~\eqref{eq:channel} separately.

\subsubsection{Path Loss and Large-Scale Fading}\label{sec:ChannelModel}
\paragraph{Air-to-Air channels:}
The simplest path-loss model assumes a \acrshort{LOS} link between the Tx and Rx and propagation in free space. This assumption represents well the situation when two UAVs communicating with each other (so-called \acrshort{A2A} channels) at a relatively high altitude (above the rooftop level). In this case, the received signal power is given as \cite{molisch2011wireless}
\begin{equation} \label{eq:Rx_FSPL}
P_R=P_T G_T G_R \Bigg(\frac{\lambda}{4\pi d}\Bigg)^2,
\end{equation}
where $P_T$ is the transmitted power, $G_T$ and $G_R$ are the transmit and receive antenna gains, respectively, $\lambda$ is the carrier wavelength, and $d$ is the distance between the Tx and Rx\footnote{For simplicity of notation, $d=d_{3d}$ in Figure~\ref{fig:A2G_param}}. 
Note that the \acrshort{PLE} $\eta$ (the power of the distance dependence) in this equation is 2 for free-space propagation. 
So that the path loss can be expressed for a generalized case as 
\begin{equation} \label{eq:FSPL}
\Lambda = \Bigg(\frac{4\pi d}{\lambda}\Bigg)^\eta.
\end{equation}
\paragraph{Air-to-Ground channels:}
Unfortunately, the signals in real-life \acrshort{A2G} wireless communications do not experience free space propagation. In the majority of literature, the well-known log-distance PL model  with free-space propagation reference is used for \acrshort{PL} (in dB) modeling:
\begin{equation}\label{eq:PL_log}
\Lambda(d)= \Lambda_0 + 10 \eta \log (d/d_0),
\end{equation}
where $\Lambda_0$ is the PL at reference distance $d_0$ ($\Lambda_0$ can be specified or calculated as free space PL $20\log\Big[\frac{4\pi d_0}{\lambda}\Big]$).
\par
When the deterministic path-loss is removed, mean power, averaged over about 10-40 wavelengths, itself shows fluctuations over time. These random variations of locally averaged received power over large distances, typically on the order of a few tens or hundreds wavelengths, are known as \acrshort{LS}, due to large obstacles such as buildings,  vegetation, vehicles, the \acrshort{UAV}'s airframe etc.  
The obstacles affecting the propagation of radio signal can be very different from each other, resulting in large-scale variations at different locations, while having approximately the same Tx-Rx distance. 
At any distance $d$, LS fading $X_{LS}$ measured in dB is usually modeled as a normal random variable with a variance $\sigma$, which takes into account random variations of the received power around the path loss curve.
This model is attractive since it is widely used (with different parameters) for modeling of classical terrestrial channels.
\par
Another common \acrshort{PL} model used in the literature \cite{3gpp_uav,6863654,azari2017ultra,azari2016joint,8038869,mozaffari2016efficient}, separates the path loss into two components namely \acrshort{LOS} and \acrshort{NLOS}:
\begin{equation}\label{eq:PLavg}
	\Lambda_{avg}=P_{LOS}\cdot \Lambda_{LOS} + (1-P_{LOS})\cdot \Lambda_{NLOS},
\end{equation}
where $\Lambda_{LOS,NLOS}$ are the path loss for the LOS and NLOS cases, respectively, $P_{LOS}$ denotes the probability of having a LOS link between the UAV and the ground node.
An advantage of this model is that $P_{LOS}$ calculation can be adapted to different heights of the communicating terminals so that we can take into account the difference between the scenarios when the aerial node communicates with a node located at low altitudes ($\leq 2$~m) or with a \acrshort{BS}, which typically are deployed higher.
\subsubsection{Small-Scale Fading}
Small-scale fading describes the random fluctuations of the received power over short distances, typically a few wavelengths, due to constructive or destructive interference of \acrshort{MPC}s impinging at the receiver. 
Different distributions are proposed to characterize the random fading behavior of the signal envelope, suitable for different wireless systems and propagation environments. 
The Rayleigh and Rice distributions, both based on a complex Gaussian distribution, are the most commonly used models. 
Considering a large number of \acrshort{MPC}s with amplitudes and random phases, the signal envelope of small-scale fading thus follows a Rayleigh distribution \cite{molisch2011wireless}. For \acrshort{A2A} and \acrshort{A2G} channels, where the impact of LOS propagation is high, the Ricean distribution~\cite{molisch2011wireless} provides a better fit.
\par
Small-scale fading models apply to narrow-band channels or taps in tapped delay line wideband models. 
Due to the stochastic nature of these signal variations, fading is usually modeled using statistical approaches and its models are obtained through measurements or through geometric analysis and simulations. The most popular type of small-scale models is Geometry-Based Stochastic Channel Models~(GBSCM)~\cite{Oestges-Clerckx}.

\subsection{Important Results}
Now, let us present the parameters of the most popular models so that one can choose the appropriate modeling approach. 
\subsubsection{Path Loss and Large-Scale Fading Modeling} \label{channelModel}
\paragraph{Log-distance models:}
\begin{table}
\centering
    \caption{Parameters of Pathloss and Large-scale fading models}\label{tab:PLE}

    \begin{tabular}{c |c|c|c|c| c	}
    \toprule
Scenario&Frequency, GHz& $\eta$&$\Lambda_0$, dB&$\sigma$, dB&Reference\\
	\midrule
\multirow{9}{50pt }{Suburban, Urban, Open field}&&2.54-3.037&21.9-34.9&2.79-5.3&\cite{7842372}\\
&&2.2-2.6&&&\cite{6162389}\\
&&2.01&&&\cite{6566747}\\
&&4.1&&5.24&\cite{1290436}\\
&&2-2.25&&&\cite{4917538}\\
&0.968&1.6&102.3&&\cite{7357682}\\
&5.06&1.9&113.9&&\cite{7357682}\\
&0.968&1.7&98.2-99.4&2.6-3.1&\cite{7835273}\\
&5.06&1.5-2&110.4-116.7&2.9-3.2&\cite{7835273}\\
\midrule
Over sea&&1.4-2.46&19-129&&\cite{5741881}\\
\midrule
Mountains&&1-1.8&96.1-123.9&2.2-3.9&\cite{7501562}\\
\midrule
Urban (LOS)&28&2.1&&3.6& \multirow{6}{17pt }{\cite{mm-model}}\\
Urban (NLOS)&28&3.4&&9.7\\
Urban (LOS)&38&1.9-2&&1.8-4.4\\
Urban (NLOS)&38&2.2-2.8&&4.1-10.8\\
Urban (LOS)&73&2&&4.2-5.2\\
Urban (NLOS)&73&3.3-3.5&&7.6-7.9\\
\bottomrule
		\end{tabular}
\end{table}
In Table \ref{tab:PLE}, the parameters of the PL log-distance model presented in \eqref{eq:PL_log} as well as the standard deviation for the Normal distribution $\mathcal{N}(0,\sigma^2)$ describing \acrshort{LS} can be found. Note that these results are valid for the frequency ranges and environments considered for the measurement campaigns used to parameterize the model.
For mmWave, atmospheric absorption and rain attenuation can also lead to a significant power loss. 
For example, at 28 GHz, the attenuations caused by atmospheric absorption and heavy rain over a distance of 200 m are about 0.012 dB and 1.4 dB as reported in \cite{4168199}.
\par
In the case when the distinction between LOS and NLOS situations is made (i.e. when~\eqref{eq:PLavg} is used), the critical point is the \acrshort{LOS} probability $P_{LOS}$ modeling. In \cite{6863654}, it is given by
\begin{equation}\label{eq:Plos}
	P_{LOS}=\prod_{n=0}^{m}\Big[1-\exp \big(-\frac{[h_{UAV} - \frac{(n+1/2)(h_{UAV}-h_G)}{m+1}]^2}{2 \Omega^2} \big)\Big],
\end{equation}
where we have $m= \text{}{}{floor} (d_h \sqrt{\varsigma \xi }-1) $ , $d_h$ is the horizontal distance between the UAV and the ground node, $h_{UAV}$ and $h_G$ are the terminal heights, $\varsigma$ is the ratio of land area covered by buildings compared to the total land area, $\xi$ is the mean number of buildings per  km$^2$, and $\Omega$  is the scale parameter of building heights distribution (assumed to follow Rayleigh distribution \cite{ITU1410}). In some cases, it is more convenient to express the LOS probability as a function of incident or elevation angle (e.g. in \cite{Azari2017CoverageMF}). These representations can be found in \cite{7037248,Azari2017CoverageMF}.

\paragraph{Ground level channel models:}
When the airspace is divided into slices, the channel is modeled in different ways depending  on the UAV altitude. 
The ground level ($1.5$~m~$<h_{UAV}\leq 10$) is the one providing the richest choice of possible channel models since the well-known models designed for conventional cellular networks can be used.
In this tutorial, let us describe just one of the options provided by 3rd Generation Partnership Project (3GPP) for macro-cell networks deployed in rural environments\footnote{expressions for other environments and micro-cell deployment can be found in \cite{3gpp_uav, 3gpp_100}} in \cite{3gpp_uav, 3gpp_100}.
\par

Again, since the LOS and NLOS cases are treated separately, the LOS probability has to be calculated. It is expressed as
\begin{equation}\label{eq:Plos_ground}
P_{LOS} = \left\{ \begin{array}{ll}
         1 & \mbox{if $d_h \leq 10~m$}\\
        \exp{\Big(-\frac{d_h-10}{1000}\Big)} & \mbox{if $10~m < d_h$}\end{array} \right.
\end{equation}
As soon as the LOS probability is known, the PL and \acrshort{LS} can be calculated. 
The altitude dependent \acrshort{LS} can be modeled as $\mathcal{N}(0,\sigma^2)$ with the parameters listed in Table~\ref{tab:LS}.
\begin{table}
\centering
\caption{Large-scale fading model parameters}\label{tab:LS}
\begin{tabular}{c	| c	}
\toprule
    $\sigma$, dB & Applicability\\
\midrule
    4&LOS, Ground level, $10~\text{m}\leq d_h \leq d_2$\\
    6&LOS, Ground level, $d_2~\text{m}\leq d_h \leq 10~\text{km}$\\
    8&NLOS, Ground level\\
    $4.2 \exp (-0.00046 h_{UAV})$&LOS, Obstructed LOS and High-altitude A2G \\
    6&NLOS, Obstructed LOS A2G\\
\bottomrule
\end{tabular}
\end{table}

It has been pointed out that the PL changes with the position of the communication nodes and can be calculated as:
\begin{equation}\label{eq:PL_ground_LOS}
\Lambda_{LOS}^G = \left\{ \begin{array}{ll}
         \Lambda_1 & \mbox{if $10~\text{m}\leq d_h \leq d_2$}\\
         \Lambda_2 & \mbox{if $d_2\leq d_h \leq 10~\text{km}$}\\\end{array} \right.,
\end{equation}
\begin{equation}\label{eq:PL_ground_NLOS}
\Lambda_{NLOS}^G = \begin{array}{ll}
         \text{max}(\Lambda_{LOS}^G,\Lambda_{NLOS}') & \mbox{for $10~\text{m}\leq d_h \leq 5\text{~km}$}\\
         \end{array},
\end{equation}
where \begin{equation*}
\begin{split}
\Lambda_1=  20 \log (40 \pi d_{3d} f_c/3)+\text{min}(0.03 h_{UAV}^{1.72},10)\cdot \log d_{3d}-\text{min}(0.044 h_{UAV}^{1.72},14.77)+0.002 d_{3d} \log h_{UAV},
\end{split}
\end{equation*} 
\begin{equation*}
\Lambda_2=\Lambda_1+40\log\frac{d_{3d}}{d_2},
\end{equation*}
\begin{equation*}
\begin{split}
\Lambda_{NLOS}'=& 161.04 - 7.1 \log W +7.5 \log h_{UAV}-(24.37-3.7(h_{UAV}/h_G)^2)\cdot \log h_G + \\
&+ (43.42-3.1 \log h_G) (\log d_{3d}-3)+ 20\log f_c - (3.2 \log (11.75 h_{UAV}))^2-4.97),
\end{split}
\end{equation*} 
\begin{equation*}
d_2 = 2\pi h_{UAV} h_G f_c/c.
\end{equation*} 
In the expressions above $f_c, h_B, W, c$ are the carrier frequency, average building height, the average street width, and the speed of light, respectively.
\paragraph{Obstructed air-to-ground channel models:}
The next air-space slice considers the UAV altitudes $10$~m~$<h_{UAV}\leq 40$~m, where the LOS probability in macro-cell networks in rural environments is calculated as in                                                                                                                                                                                                                                                                                                                                                             \cite{3gpp_uav}:
\begin{equation}\label{eq:Plos_10_40}
P_{LOS} = \left\{ \begin{array}{ll}
         1 & \mbox{if $d_h \leq d_1$}\\
        \frac{d_1}{d_h}+\exp{\Big(\frac{-d_h}{p_1}\Big)\Big(1-\frac{-d_1}{d_h}\Big)} & \mbox{if $d_1 < d_h$}\end{array} \right.,
\end{equation}
where \begin{equation*}
p_1=\text{max} (15021 \log(h_{UAV})-16053,1000),
\end{equation*}
\begin{equation*}
d_1=\text{max}(1350.8 \log(h_{UAV})-1602,18).
\end{equation*}

Next, the PL for LOS and NLOS cases can be calculated as 
\begin{equation}\label{eq:PL_10_300_LOS}
\Lambda_{LOS}^A = \text{max}(23.9-1.8\log h_{UAV},20) \log d_{3d} + 20\log\frac{40\pi f_c}{3},
\end{equation}
\begin{equation}\label{eq:PL_10_300_NLOS}
\Lambda_{NLOS}^A = \text{max}(\Lambda_{LOS}^A, -12+ (35-5.3 \log h_{UAV}) \log d_{3d}+ 20\log\frac{40\pi f_c}{3}.
\end{equation}
\paragraph{High-altitude air-to-ground channel models:}
For the $40$~m~$<h_{UAV}\leq 300$~m, the LOS probability equals 1 and PL can be calculated as in \eqref{eq:PL_10_300_LOS}.
\subsubsection{Small-Scale Fading Modeling}
As previously noted, the most common small scale fading distribution for A2G propagation is the Ricean \cite{azari2017ultra}. 
In general, A2G channels expose a higher influence of LOS-components than an average terrestrial link.
As in terrestrial channels, for the NLOS case, the Rayleigh fading distribution typically provides a better fit \cite{1290436,4917538} and of course, other distributions such as the Nakagami\cite{azari2017coexistence}, chi-squared ($\chi^2$) and non-central $\chi^2$ \cite{azari2016optimal,azari2016joint}, and Weibull distributions might also be employed. 
The family of $\chi^2$ distributions is attracting our attention since many of the distributions listed above are particular cases of it.
\par
In \cite{3gpp_uav}, several algorithms of generating small-scale fading are provided. 
One of these alternatives suggests that the small-scale model is used with a K-factor of 15 dB and all the remaining parameters are reused from the well-known terrestrial model in \cite{3gpp_100}, including the delay and angular spreads, the cross-correlations among the  large-scale parameters, the delay scaling factor, the number of clusters, the cluster delay and angular spreads, etc.
This is the simplest of the suggested algorithms, the other alternative can be found in \cite{3gpp_uav}.
\par
Note that the \acrshort{PL}, \acrshort{LS} and \acrshort{SS} models and their parameters for urban and micro-cell scenarios can be found in \cite{3gpp_uav, 3gpp_100}.

\subsubsection{Conclusions}
The choice of an adequate channel model depends on the targeted result.
When an approximate result is needed for a large set of areas, it is practical to apply a simple channel model that will reproduce the general propagation trends.
For instance, the log-distance model with a fixed \acrshort{PLE} is an appropriate choice.
However, in the case when a more specific environment is to be investigated, a more complex channel model might be necessary.
The most complete air-to-ground channel models consider:
\begin{itemize}
\item Path loss, large- and small-scale fading mechanisms,
\item Propagation slice (ground, obstructed A2G, high-altitude A2G),
\item Different environment types (urban, suburban, rural, open),
\item Separate parameterization of \acrshort{LOS} and \acrshort{NLOS} models.
\end{itemize}
In this section we presented the whole spectrum of statistical channel models that can be applied depending on the final goal.
\section{Aerial User Equipment Communication Performance}\label{Sec:UE}
In this Section, we detail the state-of-the-art results in the communication performance analysis for UAV networks, in scenarios where the UAV is considered to be a UE or mobile terminal. 
We start with an overview of the theoretical state-of-the-art, mainly concentrating on the the analytical works using the channel models described in Section~\ref{Sec:2}, or more precisely we develop our performance analysis framework based on the channel model separating LOS and NLOS propagation cases.
Next, we present the performance estimation of an LTE cellular network serving an UAV based on a simulator consisting of a realistic 3-dimensional urban environment combined with semi-deterministic channel models.
We proceed by giving an overview of all relevant measurement campaigns detailing the currently achieved UAV communication for existing communication technologies such as LTE and Wi-Fi.
\subsection{Theoretical Performance Analysis}\label{Sec:AUE_theor}
The feasibility of LTE-based UAV communication is often examined via field trials and simulations.
Such a focus on the experimental studies results in a highly fragmented picture of the issues related to the cellular-based communication with drones.
Surprisingly, the analytical investigation of \acrshort{AUE} scenarios is not widely studied in literature.
We believe that the theoretical frameworks to analyze the coexistence of the aerial and ground users \cite{azari2017coexistence} and the coverage probability \cite{azari2018reshaping} are necessary.
In \cite{azari2018cellular}, we presented a generic model to analyze the performance of UAVs served by a conventional cellular network.
In this tutorial, for simplicity reasons, we provide the results only for some specific cases that can be interesting from the practical deployment point of view.
Following characteristics are considered: i) coverage probability, ii) achievable channel capacity, and iii) area spectral efficiency.

Here we aim to address theoretically the following important questions:
\begin{itemize}
\item are the current and future cellular networks capable of providing adequate quality of service for \acrshort{AUE}s?
\item what are the major factors that may limit the network performance for \acrshort{AUE}s?
\item how does the flexibility of UAV design help to achieve better performance?
\end{itemize}

\subsubsection{Network Architecture}
\begin{figure}
\centering
  \includegraphics[width=.5\linewidth]{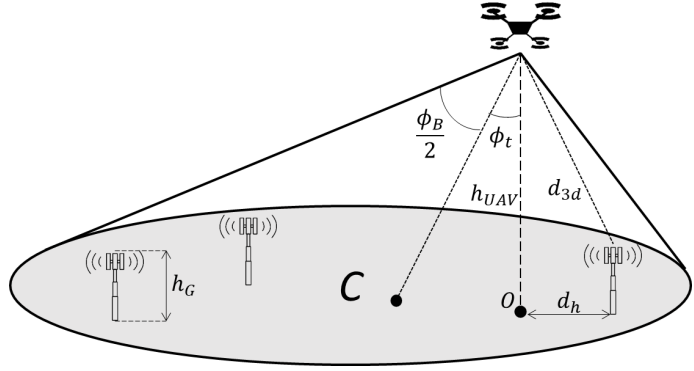}
  \caption{Network geometry}
  \label{sysmodel}
\end{figure}
\paragraph{System architecture:}
We consider a cellular network consisted of ground \acrshort{BS}s, \acrshort{UE}, and \acrshort{AUE}.
A homogeneous Poisson point process (HPPP) $\Phi$ with density $\lambda$~BSs/Km$^2$ was used to model to BSs  locations. 
The BSs' heights are denoted as $h_G$ as in \eqref{eq:Plos}. 

Users are assumed to be located $h_{UAV}$ meters above the ground (note that for ground users $h=$1.5~m). 
The horizontal distance $d_h$ separates a BS and a specific UE's projection on the ground $0$ (see Figure \ref{sysmodel}).
The BS antenna radiation pattern is imitating a realistic deployment (the antennas are vertically directional and horizontally omnidirectional).
The network is assumed to be optimized for the terrestrial users so that the antennas are tilted down \cite{3gpp2010}.
Consequently, the \acrshort{AUE} is assumed to receive signals from the sidelobes. 
The antenna gain of a BS is represented by $G_{BS}$, with $G_M$ and $G_m$ being the main- and side-lobe gains, respectively.
\par
We consider that \acrshort{AUE} is able to control the antenna tilt (mechanically or electrically). 
The \acrshort{AUE} antenna is characterized by its opening angle $\phi_{B}$ and tilt angle $\phi_t$, as illustrated in Figure~\ref{sysmodel}. 
We assume that the UAV antenna gain is $G_{UE} = 29000/\phi_{B}^2$ within the main lobe and zero outside of the main lobe \cite{constantine2005antenna}. 
As a result, an \acrshort{AUE} receives with sufficient gain only signals from BSs within an elliptical section, denoted by $C$ (see Figure~\ref{sysmodel}).
The communication link length $d_{3d}$ between a BS and a UE is defined as in Section~\ref{Sec:2}.  
\paragraph{Channel model:}
The approach presented in Section \ref{Sec:2}, equation \eqref{eq:PLavg} is used so that the \acrshort{LOS} and \acrshort{NLOS} components are treated separately with the probability of \acrshort{LOS} modeled as in \eqref{eq:Plos}.
Note that the LOS probability of different communication links are assumed to be independent.
When the LOS probability is known, path loss is calculated as in \eqref{eq:PL_log} with the reference loss $\Lambda_{L,N}$ and PLE $\eta_{L,N}$ for LOS and NLOS links, respectively.
\par
For modeling of the small-scale fading $X_{SS,\upsilon}$ ($\upsilon$ is chosen depending if the link is LOS or NLOS) we use the Nakagami-m distribution, which contains a wide range of fading types as specific cases \cite{rosas2013nakagami}. 
Accordingly,  $X_{SS,\upsilon}$ follows a distribution with the cumulative distribution function (CDF) 
\begin{equation}
F_{X_{SS,\upsilon}}(\omega) \triangleq \mathbb{P}[X_{SS,\upsilon} < \omega] = 1-\sum_{k=0}^{m_\upsilon-1} \frac{(m_\upsilon \omega)^k}{k!} \exp(-m_\upsilon \omega),
\end{equation}
where $m_\upsilon$ is the fading parameter assumed to be a positive integer for the sake of analytical tractability. 
Note that the larger $m_\upsilon$ corresponds to lighter fading.

If a BS transmits with a power level $P_{Tx}$, the corresponding received power is given by
\begin{equation}\label{eq:Prx}
P_{Rx}(d_{3d}) = P_{Tx}\,G\,\Lambda_\upsilon (d_{3d})\,X_{SS,\upsilon},
\end{equation}
where $G = G_{BS}\,G_{UE}$ represents the cumulative effect of transmitter and receiver antenna gains.

\paragraph{Performance metrics:}
The system performance is estimated using three metrics derived from \acrfull{SINR} expressed as
\begin{equation}\label{eq:SINR}
SINR = \frac{P_{Tx}\,G\,\Lambda_\upsilon (d_{h})\,X_{SS,\upsilon}}{I+N_0};~~\upsilon \in \{\mathrm{L},\mathrm{N}\}.
\end{equation}

The performance is affected by many system parameters including the AUE altitude, $h_{UAV}$. 
Therefore, all the metrics are dependent on these parameters, even if it is not expressed explicitly.

\textit{Coverage Probability} denoted by $P_{cov}$ reflects the \textit{reliability} of the link between a UE and its associated BS in satisfying the target requirement and it is defined as
\begin{equation}\label{eq:def_covp}
P_{cov} \triangleq \mathbb{P}[SINR>\mathrm{T}],
\end{equation}
which can be written as $P_{cov} = P_{cov}(h_{UAV},\mathrm{T})$ as SINR depends on $h_{UAV}$.
The target value $\mathrm{T}$ is determined based on the user requirement and is related to the target rate $R_{Tx}$ by $\mathrm{T} = 2^{R_{Tx}/\mathrm{BW}}-1$, where $\mathrm{BW}$ is the bandwidth allocated to each user. 
Additionally, this metric is useful to evaluate the reliability of a command and control link.

\textit{Channel Capacity} denoted by $\mathcal{R}$ is the highest bit rate achievable by a UE in the network. 
This metric can be calculated as
\begin{equation} \label{R-def}
\mathcal{R} \triangleq \mathbb{E}\big[\log_2(1+SINR)\big] \quad (\mathrm{b/s/Hz}),
\end{equation}
which is the user altitude $h_{UAV}$ and the BS density $\lambda$ dependent, and hence can be written as $\mathcal{R} = \mathcal{R}(h_{UAV},\lambda)$.
While the coverage is useful to characterize the quality of service and reliability, the throughput $\mathcal{R}$ quantifies performance in terms of average raw throughput.
In other words, $\mathcal{R}$ and $P_{cov}$ are complementary for the BS to UE link quality estimation.

\textit{Area Spectral Efficiency (ASE)} denoted by $\mathcal{A}$, is a network-level performance metric that reflects the achievable data-rate per square meter. 
Let us denote the ratio of UE to AUE numbers as $\rho$ . 
In a given area, the densities of BSs serving AUEs and ground users are $\lambda\rho$ and $\lambda(1-\rho)$, respectively. 
Therefore, the average throughput per square meter can be calculated as
\begin{equation} \label{ASE-def}
\mathcal{A} \triangleq \lambda [ (1-\rho) \cdot \mathcal{R}(1.5,\lambda)+ \rho \cdot \mathcal{R}(h_{UAV},\lambda)] \quad (\mathrm{b/s/Hz/\mathrm{Km}^2}),
\end{equation}
where $\mathcal{R}(1.5,\lambda)$ and $\mathcal{R}(h_{UAV},\lambda)$ corresponds to ground users at altitude of 1.5\,m and aerial users at altitude of $h_{UAV}$, respectively. 
This metric provides an insight into the overall effect of adding aerial users in the network and how the overall spectrum efficiency changes when network resources are shared between ground and aerial users.

\subsubsection{Important Results}
\paragraph{Performance analysis:}
Below, we present several results for the performance metrics described above.
Note that this tutorial contains only the expressions for the most popular practical cases (i.e. omnidirectional AUE antenna is considered), for more generalized results refer to \cite{azari2018cellular}.
\par
\textit{Coverage Probability} - First, we present the expression for the downlink coverage probability of a cellular-connected UAV equipped with an omnidirectional antenna in the case when the \acrshort{AUE} is flying higher than the serving BS.
Given the system model and performance metrics defined earlier, it is obtained as
\small
\begin{align} \nonumber
P_{cov}  \approx 2 \int_{0}^{r_e+r_M} {P}_{{cov}~|~d_h}(d')~f({d'})[\pi+\varphi_1(d')-\varphi_2(d')]~ d' \mathrm{d} d',
\end{align}
\normalsize
where $ f({d'}) \approx \lambda {P} (d') \cdot e^{-2\lambda \mathcal{I}_{1\mathrm{L}}^\mathrm{L} }$ is the probability density function (PDF) of the serving BS's distance $d_h$ at an arbitrary angular coordinate within $C$,
\begin{align} \label{pcovrl-app1}
{P}_{{cov}~|~d_h} \approx \sum_{k=0}^{m_\mathrm{L}-1} \frac{(-y_\mathrm{L})^k}{k!} \cdot \frac{d^k}{d y_\mathrm{L}^k} e^{-2 \lambda \mathcal{I}_{2{\mathrm{L}}}^\mathrm{L}},
\end{align}
\begin{align} \nonumber
y_\mathrm{L}=\frac{m_L T}{P_{Tx} G \Lambda (d_h)},
\end{align}
\begin{align} \nonumber
\mathcal{I}_{2L}^{L} \triangleq \int_{d_h}^{\infty} d' P(d') \pi&\, \Bigg[1-\left(\frac{m_L}{m_L+y_\mathrm{L} P_{Tx} G \Lambda(d_h)}\right)^{m_L}\Bigg]  \mathrm{d}d'
\end{align}

\textit{Channel capacity and ASE} - The achievable throughput of a typical UE can be obtained as
\begin{align}
\mathcal{R} \triangleq \mathbb{E}\big[\log_2(1+SINR)\big]  = \frac{1}{\ln{2}} \int_0^\infty \,\frac{P_{cov}(h_{UAV},t)}{1+t} \,\mathrm{d}t \approx \frac{1}{\ln{2}} \sum_{n = 1}^{\mathrm{K}} \frac{P_{cov}(h_{UAV},t_n)}{1+t_n} \cdot  \frac{\pi^2\sin\left(\frac{2n-1}{2\mathrm{K}}\pi\right)}{4\mathrm{K}\cos^2\left[\frac{\pi}{4}\cos\left(\frac{2n-1}{2\mathrm{K}}\pi\right)+\frac{\pi}{4}\right]}, \label{capacity}
\end{align}
where the last equation is an approximation to facilitate numerical calculations. Note that parameter $K$ has to be chosen large enough for a high accuracy of the approximation~\cite{6152071}. Also, $t_n$ stands for
\begin{align}
t_n &= \tan\left[\frac{\pi}{4}\cos\left(\frac{2n-1}{2\mathrm{K}}\pi\right)+\frac{\pi}{4}\right].
\end{align}

Finally, ASE is obtained by a direct substitution of $\mathcal{R}(h_{UAV},\lambda)$ from \eqref{capacity} into \eqref{ASE-def}.

\begin{figure}
\centering
     \includegraphics  [width=.5\linewidth] {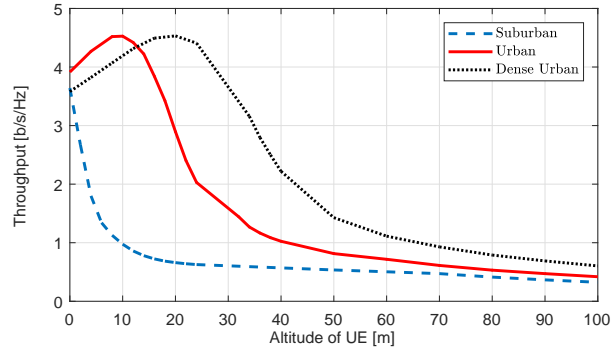} 
    \caption{The impact of UAV altitude and environment on the performance of network.}
    \label{Rate_hD_env}
\end{figure}

\paragraph{Representative case-studies:}
Considering a typical UAV served by the ground cellular network in downlink, the following results can be observed:

\textit{Altitude Impact} - Figure \ref{Rate_hD_env} shows that for the scenario when an omnidirectional antenna is used at the AUE,  the performance of the network at relatively high altitude is very low.
This is due to the growing number of BSs in \acrshort{LOS} seen by the UAV when increasing its altitude \cite{azari2018reshaping}. 
However, there is an optimum altitude at which the performance of network is maximized. 
The existence of this optimal altitude can be explained by the fact that the serving BS has a higher probability of being in \acrshort{LOS} with the AUE, while the probability of LOS on the interfering links from the other BSs is still lower due to the longer horizontal distance (see \eqref{eq:Plos}).  

\textit{Environment Impact} - Figure \ref{Rate_hD_env} illustrates the effect of different type of urban areas. When the area is more dense, the \acrshort{AUE} at relatively high altitudes benefit from more interference blocking and hence its performance is higher. As can be seen, the optimum altitude is also higher for more obstructed areas. For a suburban area, the UAV should fly as low as possible for better performance.

\begin{figure}
\centering
     \includegraphics  [width=.5\linewidth] {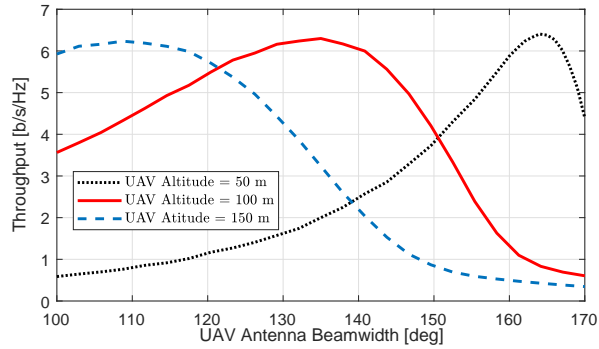} 
    \caption{The impact of UAV antenna beamwidth on the performance of network.}
    \label{Rate_fB_hD}
\end{figure}

\begin{figure}
\centering
     \includegraphics  [width=.5\linewidth] {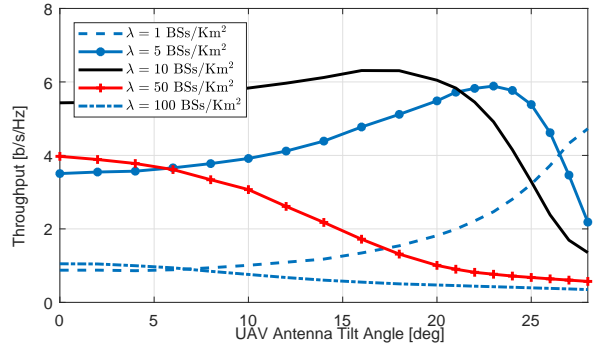} 
    \caption{The impact of UAV antenna tilt angle on the performance of network.}
    \label{Rate_fs_lam}
\end{figure}

\textit{UAV Antenna Configuration} - As it was mentioned above, the low performance of the network for aerial UEs can be compensated by using a tilted directional antenna on the UAV. In this manner, the UAV can attenuate the signals coming from the interfering BSs and hence boost the SINR levels. Figures \ref{Rate_fB_hD} and \ref{Rate_fs_lam} demonstrate the potential benefits of the optimum antenna configuration in terms of beamwidth and tilt angle. Interestingly, the performance of the \acrshort{AUE} at the optimum point is even higher than a ground UE. The optimum angle depends on the altitude of flying UAV and density of the network. As can be seen, tilting the UAV antenna is not beneficial for very dense networks since the number of the interferers is too high. 

\begin{figure}
\centering
     \includegraphics  [width=.5\linewidth] {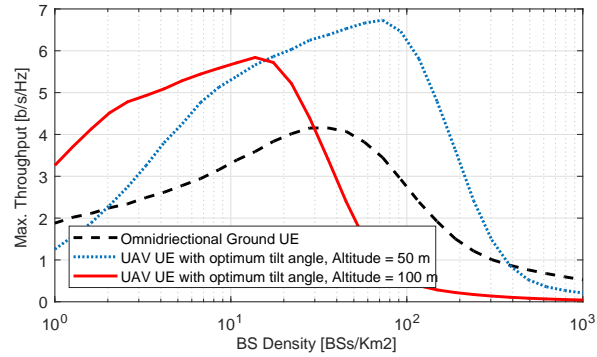} 
    \caption{The impact of network density on the UEs performance.}
    \label{Rate_lam_hD}
\end{figure}

\textit{Network Densification -} As the network becomes more dense, the performance of UEs first increases due to higher probability of LOS with the serving BS. However, further densification causes the performance degradation due to a higher number of interfering BSs. As can be seen, an AUE is capable of achieving higher performance when its antenna is configured optimally. In any case, for ultra dense networks the performance converges to zero. Finally, from Figure \ref{Rate_lam_hD} we note that, in order to mitigate the interference effect and increase the performance, a UAV flying altitude should be lower when the network density increases.

A more detailed discussion on this topic can be found in \cite{azari2017coexistence,azari2018reshaping,azari2018cellular}.

\subsection{Simulation of LTE Networks Performance}\label{Sec:AUE_sim}
In order to study the interference influence on the network performance, we designed a semi-deterministic downlink simulator combining multi-band channel models with a 3D map of a real Belgian city. Table \ref{tab:param} contains the main simulation parameters.
\subsubsection{Simulator}
\paragraph{Environment:}
\begin{table}[]
\caption{Simulation parameters}
\centering
\begin{tabular}{c|c}
\toprule
\multicolumn{1}{c|}{\textbf{Parameters}} & \textbf{Values}    \\ \midrule
Carrier frequencies                       & 1.8~GHz    \\
Signal bandwidth, B                          & 20~MHz             \\
White noise power density, $N_0$ \cite{3gpp-rf-scenarios} & -174~dBm/Hz \\ 
UE noise figure, F \cite{3gpp-rf-scenarios} & 9~dB          \\
Noise power, N                             & -122~dB        \\
Active user density, $\rho$                & 20~users/$km^2$ \\
\# iterations                             & 100                \\
 \bottomrule
\end{tabular}
\label{tab:param}
\end{table}

\begin{figure}[t]
    \centering
    \includegraphics[width=.5\linewidth]{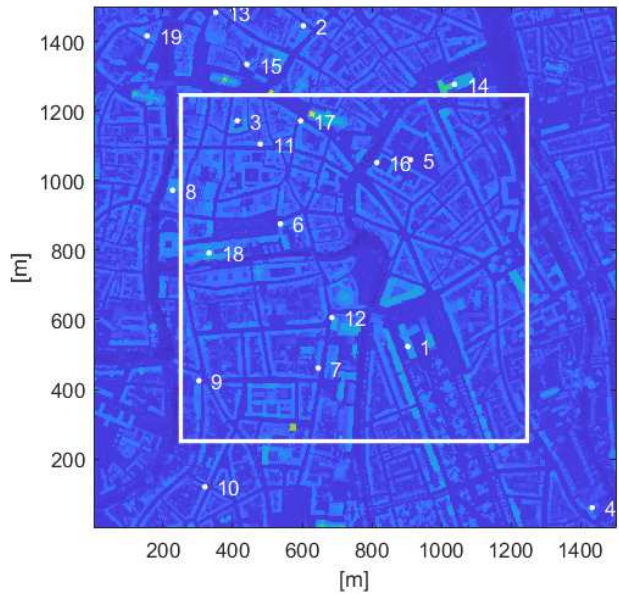}
    \caption{Surface map of Ghent city considered for the simulations}
    \label{fig:surface_map}
\end{figure}

For all simulations, a measured urban 3D environment was used. 
The 3D surface of Flanders, Belgium \cite{agiv} scanned with 1~m resolution was used. Measured data contains the height map including buildings as well as vegetation (see Figure~\ref{fig:surface_map}).  
A typical European middle-size city (Ghent) was chosen for the simulations. 
The environment is categorized as urban.
An area of 1~km$^2$ (see Figure~\ref{fig:surface_map}) centered at N$\ang{51}2'57''$ E$\ang{03}43'41''$ is used for the analysis.
\par
Real locations of the macro \acrshort{BS} of a single operator provided by \cite{bipt} were used. 
19 BSs in a radius of 750~m around the center of the map were considered. 
To imitate the mast deployment, the height of the BS  was chosen to be 5~m higher than the roof where it was deployed.
The BS heights range from approximately 23~m to 32~m, with the average height being 27.81~m. Every base station had 3 sectors.

Figure \ref{fig:surface_map} shows the map with the region of interest marked by the white lines and the base station locations marked as white dots.

\par

\paragraph{Channel modeling:}
\label{sec:link_model}
Let us now describe all the components of the used link model.
\par
\textit{Antenna patterns} - The BS antennas are modeled following \cite{itu-antenna}. 
The 3~dB beamwidth ($\phi_3$), maximum transmit power ($P_{tx}$) and maximimum antenna gain ($G_0$) values were provided by \cite{bipt}. 
An electrical downtilt of 8 degrees and no mechanical tilt is used throughout all simulations. 
The first sector is pointed north, 0~deg, and the other sectors are all evenly spaced to other angles. 
An omnidirectional antenna with 2.15~dBi maximum gain is used at both ground \acrshort{UE} and \acrshort{AUE} terminals.
\par
\textit{Line-of-sight} - To check whether the (A)UE is in \acrshort{LOS} with a BS, a line is drawn between the basestation and the user in the 3D environment. 
Analyzing the intersections with the 3D environment the link between user and basestation is considered LOS or \acrshort{NLOS}.
\par
\textit{SINR calculation} - For the \acrshort{SINR} calculations, all sectors from all BSs are assumed to be transmitting. 
Thus the received power from every sector is calculated as:
\begin{equation}
P_{rx}=P_{tx}+G_{tx}+G_{rx}-\Lambda,
\end{equation}
where $P_{rx}$ is the received power, $P_{tx}$ is the constant transmitted power, $G_{tx}$ and $G_{rx}$ are respectively the transmitting and receiving antenna gains in the direction of the user and $\Lambda$ is the corresponding path loss for that location, with all parameters in dB. 
\par
\begin{figure*}[t]
  \centering
  \subfigure[Ground user cell assignment]{\includegraphics[width=0.45\linewidth]{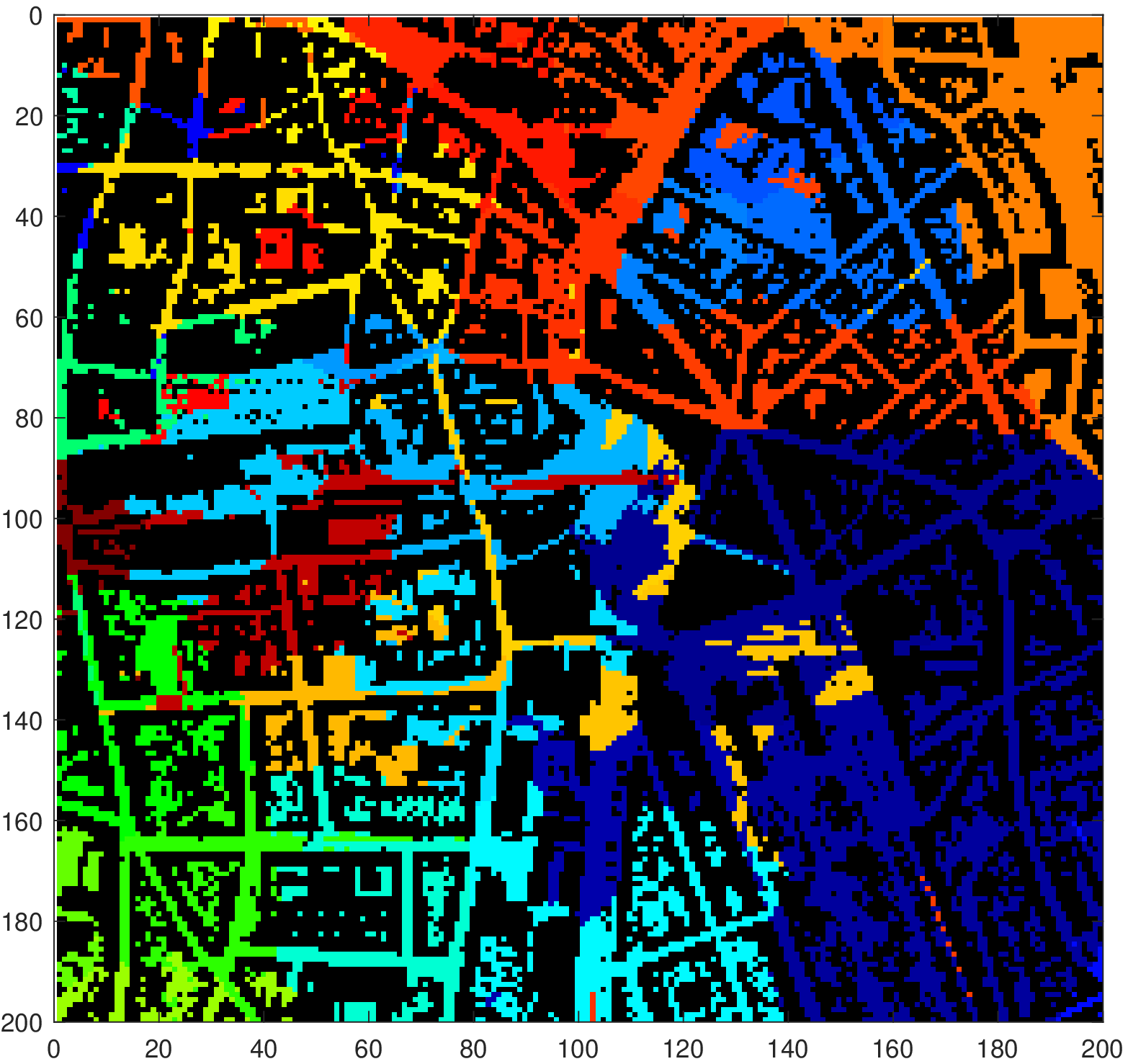}}\quad
  \subfigure[UAV user cell assignment at 150m]{\includegraphics[width=0.45\linewidth]{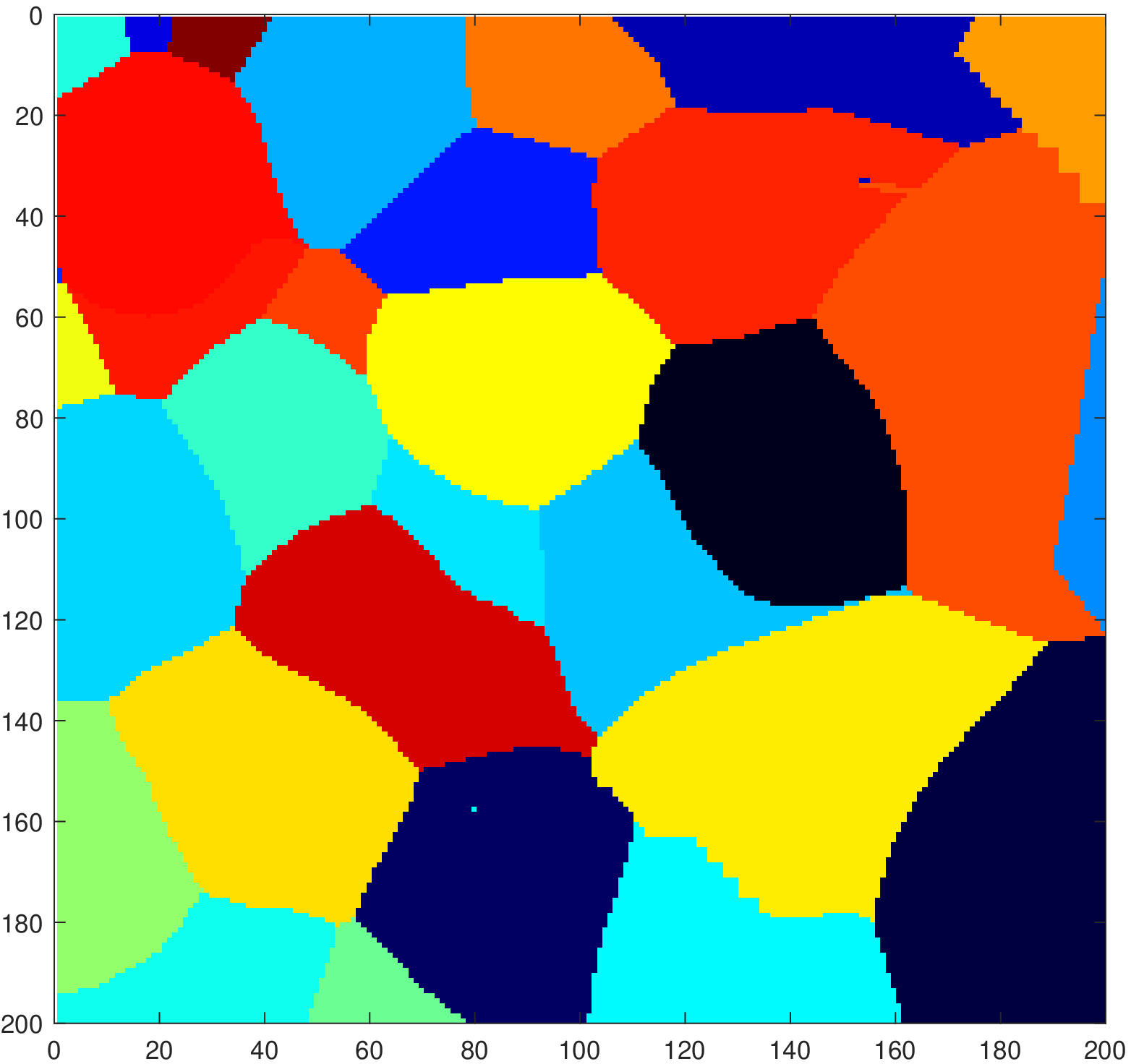}}
  \caption{Cell assignment at different heights}
  \label{fig:assignments}
\end{figure*}

Let A be the collection of all sectors. SINR levels are calculated for a given location as follows:
\begin{equation}
\forall a\in A: SINR_{a} = \frac{P_{rx, a}}{I_a + N},
\end{equation}
where $N = N_0BF$  ($N_0$ is white noise density, B is signal bandwidth and F is noise figure) is the noise power (values in Table \ref{tab:param}) and $I_a$ is the interference power considering $a$ is the serving sector, which is defined as follows:
\begin{equation}
I_a = \sum_{i\neq a}^{A}{P_{rx,i}}.
\end{equation}
The serving sector for every location is thus chosen as the one with the highest SINR value. 
After performing the simulations taking into account all parameters described above, a 3D sector assignment map that shows the sector assignment in a geographical area is generated. 
Two slices of this assignment map at different heights are shown in Figure \ref{fig:assignments}.
\par
\textit{Coverage} - The link performance is estimated using the coverage probability, which is the probability that a target SINR is achieved (as function of UAV altitude) similar to Section~\ref{Sec:AUE_theor}.
As it was mentioned above, the target SINR depends on the throughput requirement. In the case of command and control downlink, this results in an estimated data rate of only 60-100~kbps for the downlink \cite{3gpp-summary}. In \cite{3gpp-summary}, it was shown that  a minimum SINR of -6~dB is enough for this purpose.

\subsubsection{Simulation Results}
\paragraph{Environment characterization:}
First of all, some characterization and analysis of the environment is done. Basic parameters such as building height and LOS probability are calculated for later use in the channel model. 
\par
\begin{figure}[t]
    \centering
    \includegraphics[width=.5\linewidth]{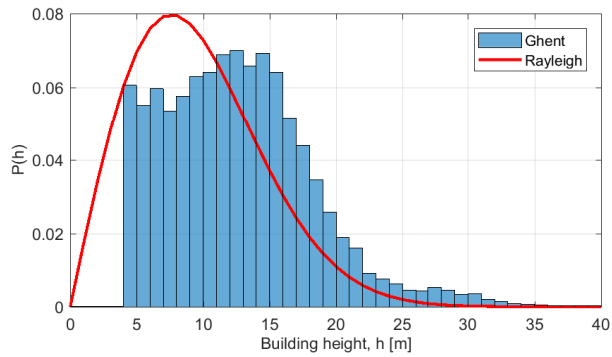}
    \caption{Distribution of building height in Ghent compared to a Rayleigh distribution}
    \label{fig:building_height}
\end{figure}

A comparison between the resulting building height distribution and the  Rayleigh distribution suggested in \cite{ITU1410} can be seen in Figure \ref{fig:building_height}. 
It turns out that a minimum building height of 4~m. As can be seen, the distribution overestimates the amount of building with a height around 10~m. The calculated mean building height is 13.32~m.
\par
\begin{figure}[t]
    \centering
    \includegraphics[width=.5\linewidth]{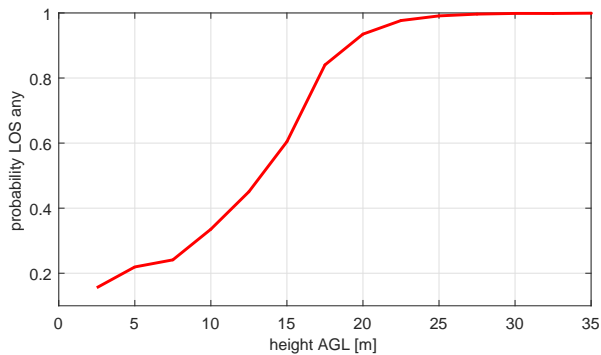}
    \caption{Probability of line of sight with at least one basestation depending on height}
    \label{fig:line_of_sight}
\end{figure}

The LOS probability is also calculated to characterize the environment: it is presented as a function of AUE height and indicates whether the users is in line of sight with any basestation. 
When determining whether a ground or aerial user is in LOS with a basestation the 3D map is used, as explained in Section \ref{sec:link_model}. 
The resulting line of sight probability can be seen in figure \ref{fig:line_of_sight}. 
Around a height of 30~m AGL a UAV is a 100\% certain to be in LOS of at least one basestation. 

\begin{figure}[t]
    \centering
    \includegraphics[width=.5\linewidth]{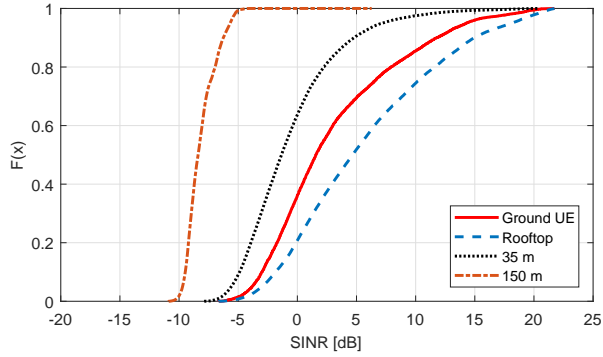}
    \caption{Cumulative Distribution Function of SINR levels for different AUE altitudes}
    \label{fig:full}
\end{figure}

\begin{figure}[t]
    \centering
    \includegraphics[width=.5\linewidth]{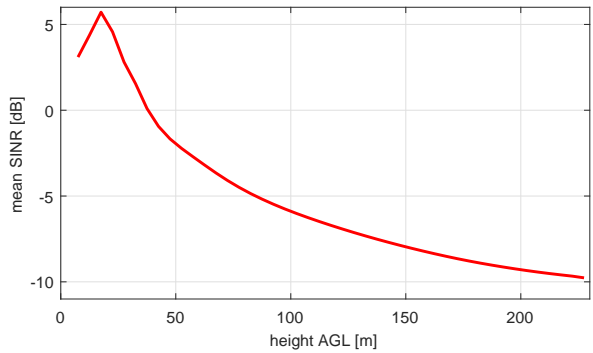}
    \caption{Mean SINR for different AUE altitudes}
    \label{fig:mean_sinr}
\end{figure}

\paragraph{Performance estimation:}
First, let us analyze how the SINR levels change with the flight altitude. 
Figure~\ref{fig:mean_sinr} shows the altitude dependent behavior of the mean SINR:
it is significantly lower for high UAV altitudes because of the interference.
A peak is observed for the rooftop level.
The mean SINR is positive for the heights lower than 40~m, however, the performance is rather defined by the SINR distributions.
The cumulative distribution function (CDF) shown in Figure~\ref{fig:full} confirms that SINR drops for high UAV altitudes, and the best performance is achieved at rooftop level.
The possible explanation of this behavior is given by the fact that the probability of having a LOS connection with the serving basestation is high at the rooftop level, while all interfering eNbs, which are at a larger distance from the user, will still be in NLOS. 
This finding is confirming the results presented in Section~\ref{Sec:AUE_theor}. 
\par
\begin{figure}[t]
    \centering
    \includegraphics[width=.5\linewidth]{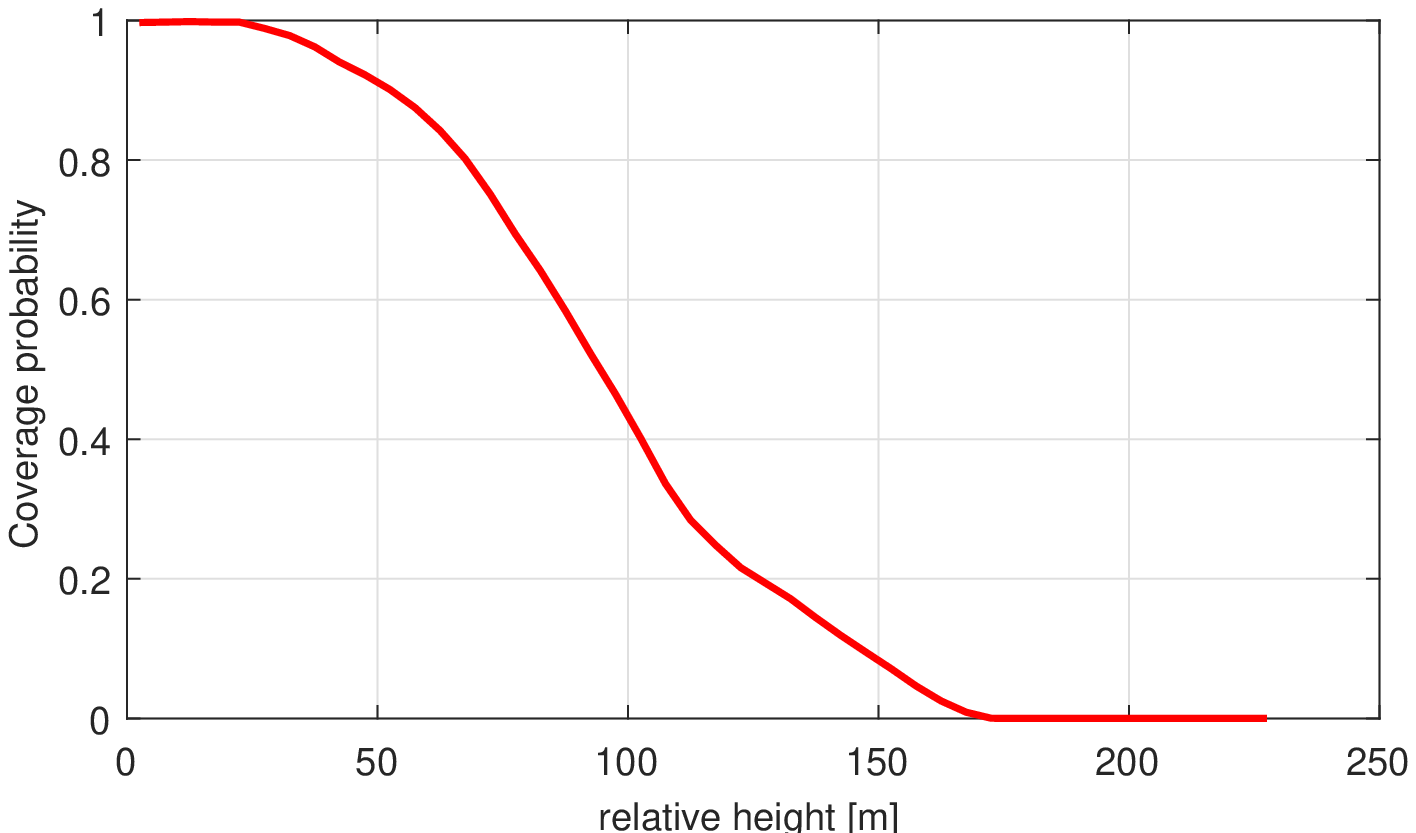}
    \caption{Coverage probability vs AUE altitudes}
    \label{fig:full_coverage}
\end{figure}

By analyzing the SINR distribution, we defined the coverage probability as a function of the \acrshort{AUE} altitude as presented in Figure~\ref{fig:full_coverage}.
It turns out that the current network configuration allows to use LTE for the UAV command and control link only for relatively low heights (up to 30-35~m) of the flying terminal.

\subsection{Measurement Based Performance Estimation}
The theoretical and simulation-based analysis given above help to investigate the potential performance achievable by cellular networks. 
However, it is hard to predict how the existing technologies perform in real-life network deployments due to several factors. 
The most important ones are: i) the difference between terrestrial and \acrshort{A2G} channels discussed in Section~\ref{Sec:2}; ii) often the real network configuration is not available.
\cite{7317490} has given an overview of the the implementation of existing wireless technologies to the design of aerial network, and discussed the pros and cons of using each of these technologies. 

\par
First of all, the understanding of the communication requirements is necessary.
Next, we can decide which technology (or mix of technologies) can suit the communication requirements. 
Let us define these requirements. For example, for collision avoidance, it is important to have a reliable communication with a moderate amount of information to be exchanged.
If the purpose of the mission is to stream high quality video (or photos) which is a very common use of UAV nowadays, then the bitrate requirements are much higher, but losing a packet here and there is not a real issue. 
Following requirements are found in literature:
\begin{itemize}
\item Command and Control: 4.8~kbaud, reliable link with delay of 50-100~ms \cite{Schoenung}.
\item Image transfer: 1~Mbps, reliable link with delay of 50-100~ms \cite{4840581}.
\item Video streaming: 2~Mbps,reliable link with delay of 50-100~ms \cite{5534937}.
\end{itemize}
\subsubsection{Wi-Fi Connected Drones}
\begin{figure}
\centering    
    \begin{subfigure}\centering  
    \includegraphics[width=.5\linewidth]{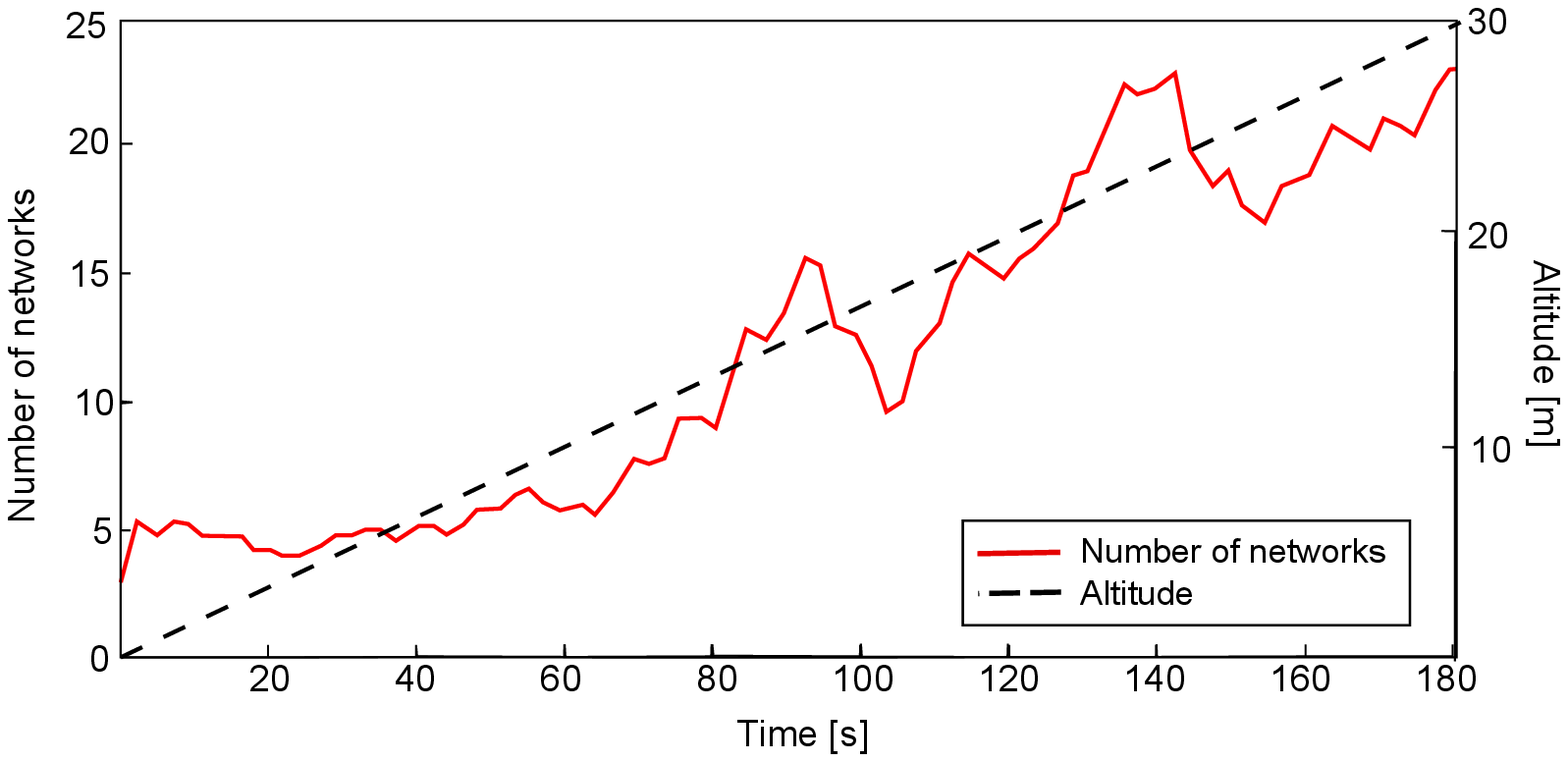}
    \end{subfigure}\\
    \begin{subfigure}   \centering  
    \includegraphics[width=.5\linewidth]{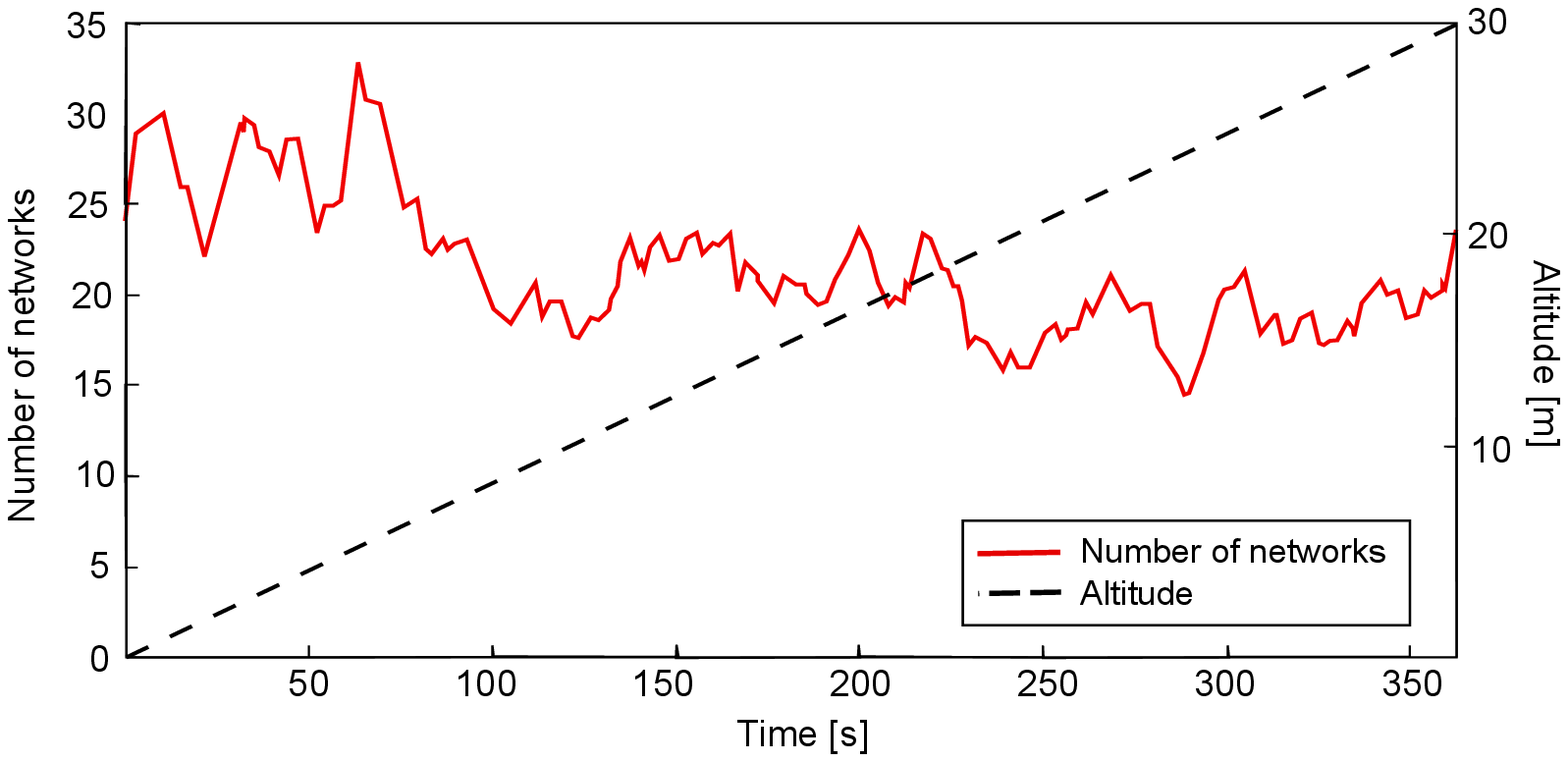}
    \end{subfigure}\\
    \begin{subfigure} \centering
        \includegraphics[width=.5\linewidth]{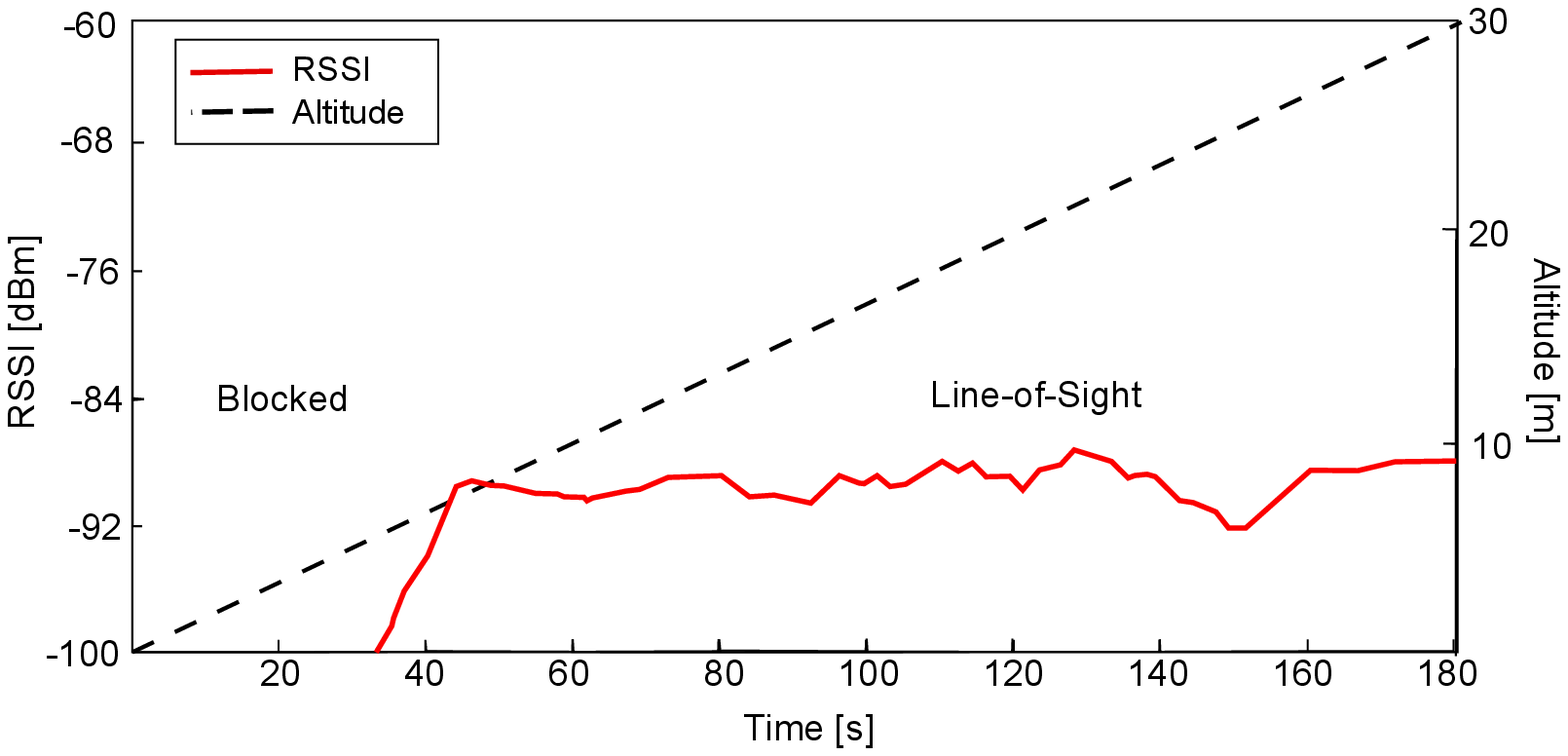}
    \end{subfigure}
    \caption{Altitude-dependent top: The number of detected networks for open space; middle: The number of detected networks for high buildings; bottom: Received signal strength of access point}
    \label{fig:wifi_neighb}
\end{figure}

A very common technology, used especially in amateur drones, is Wi-Fi. 
It is very cheap to implement because a typical smartphone or laptop can be used as remote controller. 
Practice proves that it also works quite well in Visual Line-of-Sight (VLOS) flight at very low altitude and short distances. 
However, for longer distances the changing propagation conditions must be taken into account.
Table~\ref{tab:WL_tech} presents the most valuable results obtained via various measurement campaigns for UAV communication by means of Wi-Fi (as well as Zigbee, for comparison).
It can be seen that potentially Wi-Fi is able to satisfy the requirements shown above for short distances.
\begin{table}
\centering
    \caption{Comparison of wireless technologies}\label{tab:WL_tech}

    \begin{tabular}{c |c|c}
    \toprule
    
    \multirow{2}{60pt }{Technology}&Standard &\multirow{2}{90pt }{Measured Throughput}  \\
    &PHY rate&\\
		 	\midrule
		 	Zigbee (802.15.4)&  250~kbps&up to 250 kbps (500 - 1500 m) \cite{Allred07sensorflock:an, Asadpour:2013:LDD:2535372.2535409}\\	
		 	Wi-Fi (802.11a)&  54 Mbps&UDP: 14 Mbps (350m),29 Mbps (50 m) \cite{6566747}\\
            &&{TCP: 10 Mbps (500 m), 17 Mbps (100~m) \cite{7343625}}\\
		 	Wi-Fi (802.11b)& 11~Mbps&14~Mbps (2 km)\cite{Brown2004AdHU}\\
		 	Wi-Fi (802.11n)& 600 Mbps& TCP: 10 Mbps, 100 Mbps (100 m) \cite{7343625}\\
		 	Wi-Fi (802.11ac)& 6933~Mbps& TCP: 5 Mbps (300 m), 220 Mbps (50 m) \cite{7343625}\\
		\bottomrule
		\end{tabular}
\end{table}
\paragraph{Important results:}
\begin{figure}[]
    \centering
    \includegraphics[width=.5\linewidth]{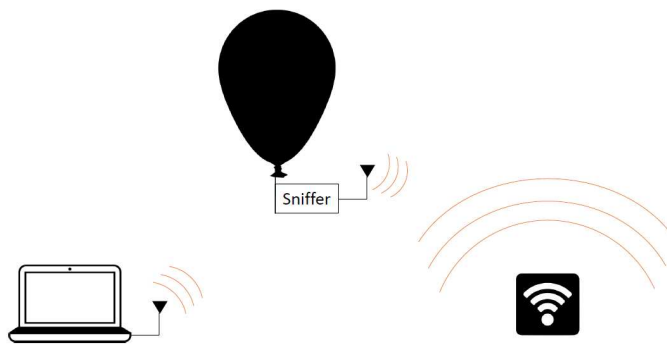}
    \caption{The wireless scanner is carried by a helium balloon.}
    \label{fig:wifi}
\end{figure}
We designed a very lightweight Wi-Fi packet sniffer that can be used on almost any UAV, including helium balloons \cite{Bergh2015AnalysisOH}. 
It was used (see Figure \ref{fig:wifi}) to log i) received signal levels (RSSI) of a Wi-Fi transmitter to see how it varies with height, and ii) track the number of Wi-Fi transmitters overheard.
Two different environments were considered: and open field and an area with high buildings (the balloon never reaches the rooftop height).
Our measurements confirm that in the high building case (see Figure~\ref{fig:wifi_neighb},~middle), the number of networks that are overheard by the sniffer does not increase. 
When the balloon rises above buildings, the number of overheard APs increases considerably. 
This confirms that increasing the impact of the LOS propagation results in a greater number of networks being detected at high altitude.

To further verify this, the received signal strength of an access point  is analyzed.
Typically, the access point is detected when the balloon and that AP are in line-of-sight. 
As seen on Figure~\ref{fig:wifi_neighb}~(bottom) the RSSI, when detected, remains almost constant. 
This is another verification that LOS and NLOS cases must be treated separately as it was stated in Section~\ref{Sec:2}.

\subsubsection{LTE Connected Drones}\label{Sec:meas_LTE}
The viability of  LTE commercial mobile networks for UAVs  was supported by Qualcomm \cite{QC_report} based on a large-scale measurement campaign.
Their results show that the signal quality of the downlink is statistically lower for  UAVs compared to the ground users: \acrshort{SINR} levels decrease up to 5 dB. 
However, the coverage outage probability (defined as \acrshort{SINR} $\leq$ -6dB) was found to be very similar for \acrshort{AUE} and \acrshort{UE}, so that it was concluded that commercial LTE networks should be able to support downlink communications requirements of initial LTE-connected drone deployment without any change. 
Note that these optimistic conclusions were not confirmed by \cite{8337920,8301389,lte_in_the_sky}.

The measurements in \cite{QC_report,8337920,8301389,lte_in_the_sky}  show that the interference is the main limiting factor for different environments.
In \cite{8337920}, the authors studied the feasibility of wireless connectivity for \acrshort{AUE} via LTE networks.
Note that the works \cite{QC_report,8337920,8301389} considered the  rural scenarios.
\paragraph{Important results:}
We have performed several measurement campaigns to quantify the interference to an LTE terminal as a function of altitude \cite{lte_in_the_sky}. 
A light aircraft was used to carry an airborne receiver at 150 m and 300 m altitude.
The position (measured with GPS) was also logged, giving the exact 3D locations of all the measurements. 
The followed trajectory was along the Belgian coast. 
The LTE signals in the 800 and 1800 MHz band have been digitized and recorded using an NI USRP X310 SDR. 
An ANT-IBAR-FMEF antenna from RF-solutions was attached to a window of the airplane. 
The collected data was analyzed using GNURadio, openLTE, and LTE Cell.
\par
For lower altitudes (0 - 120 m), a quadrotor UAV was used at KU Leuven, Belgium. 
Due to weight constraints, part of the UAV measurements were done using an off-the-shelf LTE-capable cell phone (OnePlus One) running an LTE cell tracking application. The used application was G-MoN for Android. 
Additional ground measurements (with RTL-SDR) were used as a reference.
\par
\begin{figure}[]
    \centering
    \includegraphics[width=.5\linewidth]{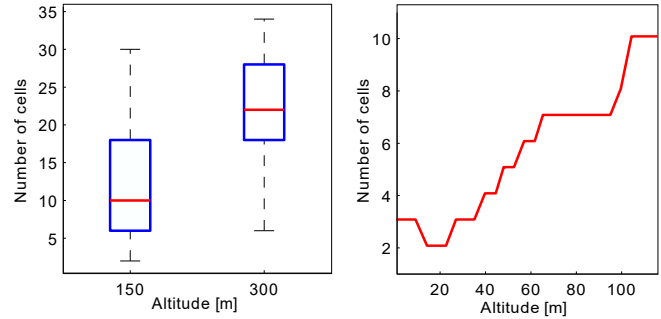}
    \caption{Number of cells seen vs. receiver altitude.}
    \label{fig:neighbours}
\end{figure}
In Figure \ref{fig:neighbours} one can see the number of cells that are visible to the receiver at different altitudes. The left plot shows the measurements performed by the UAV, while the right plot shows the results found along the coast using the airplane. To be able to compare inland and coastal measurements, taking into account that there are no base stations in the sea, the number of base stations seen at the coast has been multiplied by two.
\par
The number of the identified base stations increases significantly with altitude. 
On one hand, the airborne receiver has a high probability of being covered by the ground network. 
As a result, on the other  hand, inter-cell interference is significantly increasing  with altitude.
This leads to a decreased \acrshort{SINR} at the airborne receiver.
\par
\begin{figure}[]
    \centering
    \includegraphics[width=.5\linewidth]{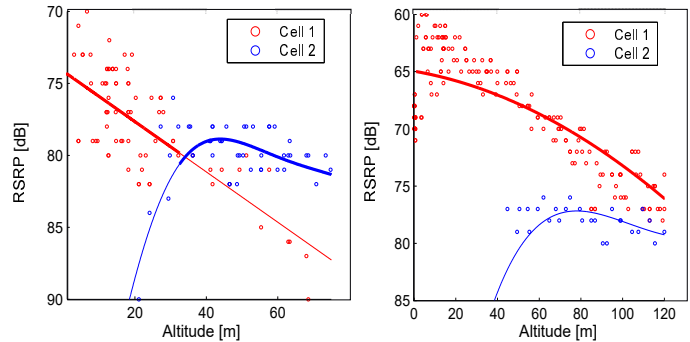}
    \caption{ The signal level as a function of the receiver altitude.}
    \label{fig:cells}
\end{figure}
Figure \ref{fig:cells} shows the reference signal received power (RSRP) signal level of the two best cells observed from a hovering UAV as a function of the altitude.
It can be seen that the signal from the best cell at ground level decreases with altitude, but signals from interfering base stations increase because of the varying propagation conditions.
Due to the fact that the networks are optimized to serve terrestrial users, the signal level from the cell that was optimal at ground level decreases. 
As the altitude increases, the signal level from the weaker cells (cell 2) increases as well due to the elimination of obstacles between the eNodeBs and the \acrshort{AUE}.
At a certain point, the attenuation caused by the obstacles is completely overcome. 
This results in a flooring of the RSRP level in Figure \ref{fig:cells}. 
\par
The measurements confirm the conclusions drawn by the theory and simulations: the downlink signal level received by the UAV from ground eNodeBs is determined by \acrshort{LOS} propagation path loss and the base station antenna gain pattern. 
\par
\par
\begin{figure}[]
    \centering
    \includegraphics[width=.5\linewidth]{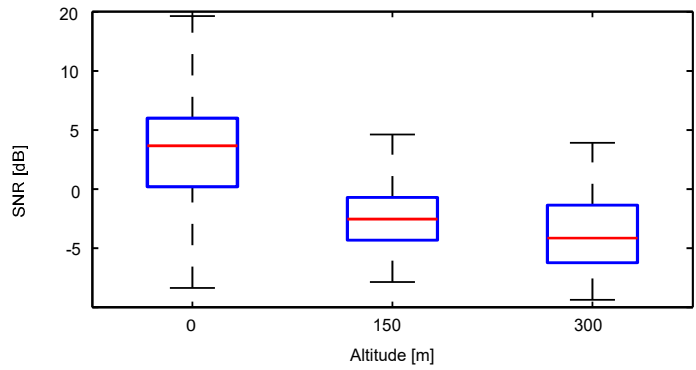}
    \caption{SINR measured on the synchronization symbols of the best cell vs.
receiver altitude (816 MHz).}
    \label{fig:sinr}
\end{figure}
Next, the SINR, measured on the synchronization symbols, is compared at three different altitudes: ground, 150 m, and 300 m, as shown in Figure \ref{fig:sinr}. 
At least 36 measurements have been taken at each altitude. 
It can be seen that the SINR of the best cell seen at each of those specific altitudes is much lower than the SINR witnessed at ground level.
There is a further slight decrease between 150 m and 300 m.
This is explained by a combination of two factors: i) increase of the distance-dependent path loss, and ii) the dramatic increase in interference levels.
While the signal strength of the best cell at ground level does go down, as shown above, the received signal form other cells might become stronger than the signal from the initial cell; thus, an UAV will handover while increasing the altitude. 
Cumulative power received from all the additional cells visible at a high altitude results in high interference.
\subsection{Conclusions}
Concluding the discussion above, we underline that the analytical approach to the AUE performance estimation is very elegant and gives the valuable information for practical network deployment. 
However, certain real-life aspects (e.g., real BS locations and buildings surrounding the BS) cannot be considered. 
On the other hand, the measurement-based performance estimation by definition takes the real environment and network configuration into account, however, naturally, this approach has several practical limitations (e.g. it cannot explore dense AUE scenarios, altitude is a limitation, and you cannot fly in all environments due to the current regulation constraints).
Simulation based approach is able to overcome the aforementioned limitations since realistic 3D maps (including the existing communication infrastructure and its parameters) are becoming available. The possible problem of simulations is that they rely on channel models. So that the appropriate channel modeling is extremely important for this approach.
\par
Summarizing this section, we would like to emphasize that all analysis (analytical, simulation and measurements) confirms that UAV communication performance is limited by interference.

\section{Unmanned Aerial Communication System Performance}\label{Sec:System}
\label{sec:UAVinfrastructure}
In this Section, we detail the state-of-the-art results in the communication performance analysis for UAV communication, in scenarios where the UAV is considered to be a part of the communication network. 
We start with an overview of the state-of-the-art, mainly concentrating on the theoretical works on optimal ABS deployment  based on using the channel models presented in Section~\ref{Sec:2}.
We proceed by a detailed description of UAV-enabled localization framework including both theoretical and simulation results.
\subsection{Theoretical Performance Analysis}
Here we aim to theoretically address the question how multiple degrees of freedom in UAVs networking (including altitude of operation, UAVs density, and their antenna configuration) impact the final performance of ABSs (coverage area, transmit power as well as the ground UE received quality of service).

\subsubsection{Network Architecture}
\begin{figure}
\centering
  \includegraphics[width=.5\linewidth]{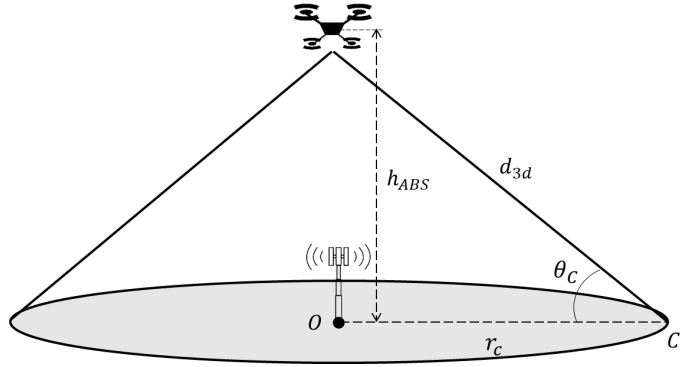}
  \caption{Network geometry}
  \label{abs:sysmodel}
\end{figure}
\paragraph{System architecture:}
We consider a hybrid air-to-ground communication network with an ABS located at an altitude $h_{ABS}$, serving ground users within its coverage area (see Figure \ref{abs:sysmodel}). 
The coverage area  is a disc of radius $r_C$ centered at $O$, the projection of the \acrshort{ABS} on the ground. 
Note that the signal arrives to ground users located at the area $C$ with elevation angles within a range $\theta_C\leq\theta\leq\pi$ depending on the distance $r$ from the center of the area, where $\theta_C=\arctan \frac{h_{ABS}}{r_C}$.

We aim to compare the performance of this network with the case when a conventional terrestrial BS is deployed at the disc center.
\paragraph{Channel model:}
In this section, for obtaining generalized results, let us consider the PL model presented in \eqref{eq:FSPL} with the \acrshort{PLE} $\eta$ that can change depending on the targeted environment and the ABS altitude.
As it was mentioned in Section~\ref{Sec:2}, the Rician distribution can be used to model the small-scale fading $X_{SS}$ between the ABS and the ground terminal.
In Section~\ref{Sec:2}, the K-factor is suggested to be equal to 15, however, it can be beneficial to have a theoretical framework considering more general dependencies, so K-factor is $\theta$-dependent.
\par
\paragraph{Performance metrics:}
The communication link between the ABS and a ground terminal is defined to be in outage when the instantaneous channel SNR falls below a given threshold $T$
\begin{equation}\label{eq:def_outp}
P_{out} \triangleq \mathbb{P}[SNR>\mathrm{T}].
\end{equation}
Consequently, the outage dependent on the communication parameters (locations of the nodes, transmit power) can be expressed as
\begin{equation}\label{eq:outp}
P_{out} (r,h_{ABS}, P_{Tx}) = 1-Q \Bigg(\sqrt{2K},\sqrt{\frac{2T [1+K] d_{3d}^\eta N_0} {G P_{Tx}}}\Bigg),
\end{equation}
where $Q(\cdot,\cdot)$ is the first order Marcum Q-function.
Note that $K,\eta$ and $G$ are $\theta$-dependent. 
Moreover, the communication distance is $d_{3d}=\sqrt{h_{ABS}^2+r^2}$.
\par
It is known that increasing the BS deployment height results in a larger coverage area, hence, an \acrshort{ABS} can either be more power-efficient for serving a fixed area or ensure coverage over a larger area transmitting the same power.
Moreover, increasing the SNR in general improves the performance of the network.
\par
\textit{Power gain} achievable by ABS in comparison with a terrestrial BS for covering the area $C$ can be defined as
\begin{equation}\label{eq:def_PG}
\mathcal{G}^P(h_{ABS}) \triangleq \frac{P_{Tx~|~BS}}{P_{Tx~|~ABS}(h_{ABS})},
\end{equation}
where $P_{Tx~|~BS}$ and $P_{Tx~|~ABS}$ are the transmit powers necessary to satisfy \eqref{eq:def_outp} when the area $C$ is served by a ground and aerial BSs, respectively.
\par
\textit{Coverage radius} characterizes the dimensions of the area within which \eqref{eq:def_outp} is satisfied. 
\par
\textit{Sum-rate gain} can be calculated when the achievable rates when terrestrial and aerial BSs are compared.
The average sum-rate of an ABS with the transmit power
$P_{tx}$ is defined as \cite{Shalmashi2016}
\begin{equation}
\mathcal{\bar{R}}(h_{ABS},P_{Tx}) = \mathcal{\bar{N}}\cdot W \log_2(1+T)[1-\bar{P}_{out} (h_{ABS}, P_{Tx})],
\end{equation}
where $\mathcal{\bar{N}}$ is the average number of ground nodes within $C$, $W$ is the transmission bandwidth, and $\bar{P}_{out} (h_{ABS}, P_{Tx})$ is the average outage probability $P_{out} (r_c, h_{ABS}, P_{Tx})$ over the coverage region $C$.
The average sum-rate gain $\mathcal{G}^\mathcal{\bar{R}}(h)$  provided by an ABS over a terrestrial one can be expressed as 
\begin{equation}\label{eq:SRG}
\mathcal{G}^\mathcal{\bar{R}}(h)=\frac{1-\bar{P}_{out} (h_{ABS}, P_{Tx~|~ABS})}{1 - \bar{P}_{out} (0, P_{Tx~|~BS})}.
\end{equation}
\subsubsection{Important Results}
\paragraph{Performance analysis:}
The transmit power required  to cover the region $C$ when an ABS is used can be obtained as 
\begin{eqnarray}\label{eq:P_Tx}
P_{Tx~|~ABS}= \frac{N_0 T}{G} \frac{2K(\theta)+2}{\Big(Q^{-1}(\sqrt{2 K(\theta)},1-\epsilon)\Big)^2}\Big[ \frac{r_c}{\cos\theta_C}\Big]^{\eta(\theta)},
\end{eqnarray}
where $\epsilon$ is the  minimal tolerable quality of the communication link, which is usually attained at the boundary of the coverage area and $Q^{-1}(\cdot,\cdot)$ is the inverse Marcum Q-function with respect to its first argument and $r_c$ is the radius of $C$.
\par
\textit{Power gain} $\mathcal{G}^P(h)$ of an ABS over a TBS can be written as
\begin{eqnarray}
\mathcal{G}^P(h)= x_0 x_\theta^{-1} r^{\eta(0)-\eta(\theta)} [\cos \theta_C]^{\eta(\theta_C)}
\end{eqnarray}
where
\begin{equation*}
x_a = \frac{2K(a)+2}{\Big(Q^{-1}(\sqrt{2 K(a)},1-\epsilon)\Big)^2}
\end{equation*}
and $a=0$ or $a=\theta$ for ground and aerial base stations, respectively.
\par
\textit{Sum-rate gain} $\mathcal{G}^\mathcal{\bar{R}}(h)$ can be calculated by replacing the transmit power from \eqref{eq:P_Tx} into \eqref{eq:SRG}.
For more results and metrics, please refer to \cite{azari2016optimal,azari2016joint,Azari2017CoverageMF,azari2017ultra}.

\paragraph{Representative case-studies:}
Considering several degrees of freedom in UAVs networking which include altitude of operation, UAVs density, and their antenna configuration, we study their impacts on UAVs coverage area and transmit power as well as the ground UE received quality of service.
\par
\begin{figure}
\centering
     \includegraphics  [width=.5\linewidth] {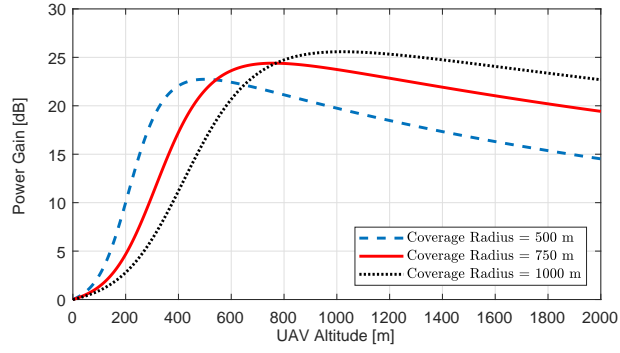} 
    \caption{Power gain of a UAV BS as compared to terrestrial BS.}
    \label{PowGain_h_Rc}
\end{figure}

\textit{Fixed Coverage Region -} Figure \ref{PowGain_h_Rc} shows how altitude of an ABS influences the required transmit power of UAV when the ABS aims to cover a certain given region on the ground. 
As can be observed, an ABS requires lower transmit power to serve a fixed area in comparison with a ground BS due to favorable LOS propagation conditions (within a certain range of altitudes). 
Particularly there is an optimum altitude at which a minimum necessary transmit power is obtained. 
This can be explained by the fact that the link is in LOS yet, at some point, the path loss becomes too high due to the increased link length.
The power gain at the optimum altitude grows as the region that needs to be covered becomes larger \cite{azari2016joint}. 
\par
\begin{figure}
\centering
     \includegraphics  [width=.5\linewidth] {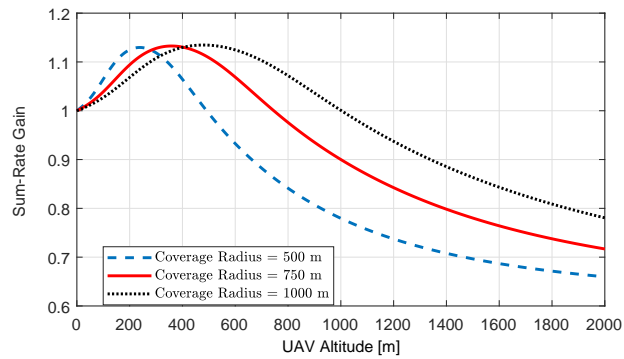} 
    \caption{Sum-Rate gain of a UAV BS network as compared to terrestrial BS.}
    \label{SumRateGain_h_Rc}
\end{figure}

Furthermore, Figure \ref{SumRateGain_h_Rc} illustrates that, at a certain range of altitudes, an ABS is able to provide higher rate capacity for ground users even though the UAV-mounted BS transmit power is lower \cite{azari2016joint}. 
However, the optimum altitude that maximizes the capacity is lower than that of power gain maximization. 
Therefore, while designing a cellular network including ABSs, the deployment altitude should be chosen carefully depending on the required performance metric.

Finally, we note that due to the absence of interference, the optimum altitudes are normally high (200 - 1000~m depending on the optimized parameter). 
However, when considering interference of neighboring ABSs, the optimum altitude might become lower \cite{Azari2017CoverageMF}.

\begin{figure}
\centering
     \includegraphics  [width=.5\linewidth] {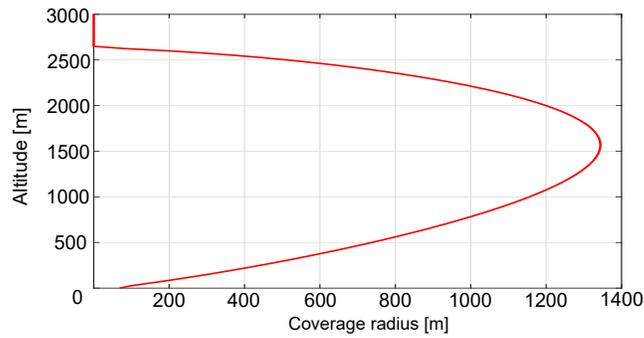} 
    \caption{Coverage radius versus altitude.}
    \label{covradius}
\end{figure}

\textit{Fixed Transmit Power -} When we consider a given transmit power, the overall effect of altitude on the coverage region is noticeable \cite{azari2016optimal}. The ABS at the optimum altitude covers significantly more users (see Figure \ref{covradius}). The trade-offs here is between a higher LOS probability at the higher altitude and longer link length. 

\begin{figure}
\centering
     \includegraphics  [width=.5\linewidth] {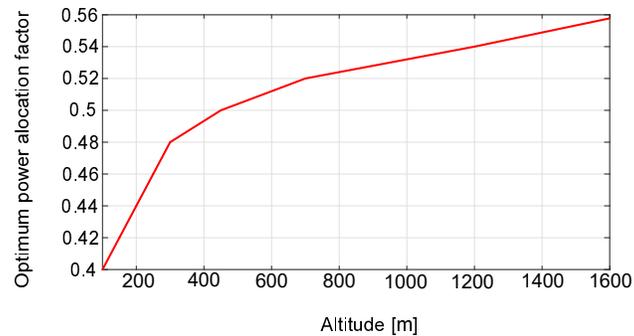} 
    \caption{Power allocation ratio versus altitude.}
    \label{OptPow}
\end{figure}

\textit{ABS Power Control -} When we consider co-channel interference caused by the neighboring BSs, the ABSs in general introduce higher interference due higher probability of having LOS communication. 
To reduce possible negative effects of this, we would like to lower the transmit power yet guaranteeing a minimum QoS for the ground users in the target region. 
As a possible solution for this issue, we suggest that cooperative communication can be employed in which a relaying ground user assists the ABS to serve another user \cite{azari2017ultra}. 
Figure \ref{OptPow} shows that by using this technique, the transmit power of UAV can be significantly lower particularly at lower altitudes. 
More discussion on this can be found in \cite{azari2017ultra}.

\textit{Multiple ABSs -} To serve a number of users distributed over a target region, multiple ABSs may be required as the coverage of a single ABS is limited. 
In this case, multiple design factors have to be taken into account. 
In a static case, in which ABSs are stationary, the inter distances of UAVs and 3D placement of them should be well defined to mitigate interference and at the same time to cover the target region \cite{mozaffari2016efficient}. 
However, when the ABSs are mobile (for instance, due to the change in the demanding spots), an appropriate UAV antenna configuration along with optimum density have to be considered. 
When distributed ABSs are considered, an appropriate UAV antenna beamwidth extends the coverage of a typical user to the point of interest. 
We can assume that there is an optimum beamwidth which maximizes the performance. 
This can be explained by the fact that the wider beamwidth is, the higher probability of detecting an ABS by a user.
However, further increasing of the beamwidth results in a higher probability of having more interfering ABSs. 
Therefore, one expects to balance these effects at an optimum value of ABSs beamwidth. 
The optimum beamwidth decreases as the altitude increases. 
\par
A similar reasoning justifies the existence of optimum density of ABSs. 
Moreover, a more obstructed urban area benefits from blocking interfering ABSs and hence the performance is higher for larger densities of ABSs. 
These discussions are more detailed in \cite{Azari2017CoverageMF,azari2018uplink}. 
Additionally, specific aspects of the uplink performance analysis of ABSs (omitted in this tutorial) are reported in \cite{azari2018uplink}. 
\subsection{Localization Services System Requirements}
A localization service is acknowledged as a fundamental functionality in modern communication systems. While it is recognized as a complementary service in ordinary scenarios, it is of critical demand in case of emergencies to locate the user and provide relief services efficiently. As all other services, localization services using ground base stations are exposed to damage caused by natural or man-made disasters. Therefore, rapid on-demand deployment of UAVs furnished with radio platforms for serving and localizing ground users has, recently, attracted  considerable research focus \cite{8255737,aerial_anchors,sallouha2018energy,perazzo,8422375}. In particular, because of the higher LOS probability at high altitudes which results in better localization accuracy compared with ground base stations \cite{aerial_anchors,sallouha2018energy}. 

\begin{figure}
\centering
     \includegraphics  [width=.5\linewidth] {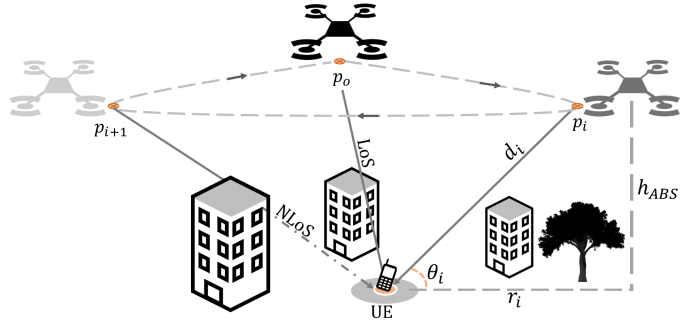} 
    \caption{An illustration of UAV(s) at different anchor points to localize a UE}
    \label{fig:locModel}
\end{figure}

The multilateration process is the most prominent method in wireless communications for accurately determining the position of the user. It substantially requires estimating the distance to the user from at least three different positions naming \textit{anchor points} as shown in Figure \ref{fig:locModel}. The distance estimation is typically done using time-based or RSS-based techniques \cite{Hazem,locUWB}. Time-based techniques estimate the distance by simply multiplying the estimated time of flight by the speed of light. However, defining the time of flight is the bottleneck of these techniques as a very accurate time synchronization is required between the transmitter and the receivers. RSS-based solutions, on the other hand, are known for their intrinsic simplicity and for not requiring any time synchronization. Moreover, RSS estimation functionality is readily available in all chipsets \cite{Zanella}. RSS-based techniques estimate the distance by using RSS-distance function which is well represented by the path loss model.

RSS-based localization techniques have been extensively studied in the literature for terrestrial cellular networks \cite{dammann2013where2,Zanella,Hazem}. 
In general, the achieved localization accuracy is limited due to the relatively high shadowing that causes a rather high distance estimation error \cite{Zanella}. To overcome the high shadowing on the ground, UAVs as aerial anchors is introduced as a novel solution to localize ground devices \cite{perazzo,aerial_anchors,Rubina}. In fact, UAV anchors are able to provide higher probability of \acrshort{LOS} with ground UE and less shadowing effect as detailed in Section \ref{sec:ChannelModel}. In the following, we address the main design parameters that affect the localization accuracy when using UAVs as an ABS. These parameters are the UAVs' altitude, trajectory radius and number of anchor points \cite{aerial_anchors,sallouha2018energy}.

For simplicity, one can assume that the UE is in the coverage range of the UAV at $i$-th anchor point for all examined values of $h_{ABS}$. Now, following the channel model presented in Section \ref{channelModel} and assuming that $\delta_{\text{LOS}}$ and $\delta_{\text{NLOS}}$ respectively represent the localization errors corresponding to LOS and NLOS components, the average localization error can be written as
\begin{eqnarray}
\delta = P_{LOS}~\delta_{\text{LoS}} + [1 - P_{LOS}] \,\delta_{\text{NLoS}}.
\label{Errtotal}
\end{eqnarray}
Without loss of generality, we assume that the UE location is $(x_g, y_g, 0)$ whereas the UAV position is $(x_a, y_a, h)$. Consequently, given the estimated distance $\hat{r}_i$ and known projection $(x_a^{(i)}\,, y_a^{(i)})$ of the UAV at the $i$th anchor point, the position of the UE can be estimated by finding the point $(\hat{x}\,, \hat{y})$ that satisfies
\begin{eqnarray}
(\hat{x}, \hat{y}) = \argmin_{x,y} \bigg\{\sum_{i=1}^{M} \Big( \sqrt{(x_a^{(i)} - x_g)^2 + (y_a^{(i)} - y_g)^2} - {\hat{r}}_i  \Big)^2 \bigg\}.
\label{opt2}
\end{eqnarray}
where $\hat{r_i} = \sqrt{\hat{d}_i - h_{ABS}^2}$, $M$ is the number of anchor points and $\hat{d}_i$ is the estimated distance estimated based on the path loss model of LOS or NLOS link. Now, for an estimated location $(\hat{x}\,, \hat{y})$ of a UE, the localization error is expressed as
\begin{eqnarray}
\delta_j = \hspace{0.1cm} \parallel {\hat{\boldsymbol{r}}} - \boldsymbol{r} \parallel \hspace{0.1cm} =  \sqrt{\sum_{i=1}^{M} \mid\hat{r}_i - r_i \mid^2}, ~~~j \in \{\mathrm{LoS}\,,\mathrm{NLoS}\}
\label{locoErr}
\end{eqnarray}
where $\boldsymbol{r} = [r_1, r_2, ..., r_M]$, ${\hat{\boldsymbol{r}}} = [\hat{r}_1, \hat{r}_2, ..., \hat{r}_M]$ and $\lVert.\lVert$ represents the euclidean distance. Estimating the distance to the UE for different anchor point can be done using single mobile UAV or multi-UAVs hovering are the two possible approaches.
\subsection{Simulation Setup}
In order to investigate the localization accuracy when using a mobile UAV, simulations have been conducted. In our simulations, we assume 100~UE uniformly distributed in a circular area with a radius of $200\,$~m. We consider a system communication frequency of 2~GHz. The relevant system parameters and their corresponding values are specified in details in Table \ref{const}.

\begin{table}[!t]
	\caption{Parameters List}
    \label{const}
	\centering
		\begin{tabular}{ c || c | c }
			\hline
			\textbf{Parameter} & \textbf{Description} &  value \\
\midrule
			$M$ & Number of anchor points &  3 , 4 \\

			$N$ &  Number of TNs &  100 \\

			$f$ & Carrier frequency [GHz]&  2  \\

			$A$ &  Total area of TNs [km$^2$] & 0.12 \\

			$\delta$ &  Localization error & -- \\

			$l_j$ &  anchor points inter distance [m] & -- \\ 

			$h_{ABS}$ & UAV's altitude & 200 \\

			$t_h$ & Hovering time [s] & 5 \\

			$R$ &  Trajectory radius & 120 \\

			$a_{\text{LoS}}$ & Shadowing constant  & 10\\

			$b_{\text{LoS}}$ &  Shadowing constant & 2 \\

			$a_{\text{NLoS}}$ & Shadowing constant & 30\\

			$b_{\text{NLoS}}$ &  Shadowing constant & 1.7 \\

			$a_o$ & $P_{LOS}$ constant  & 47 \\
			$b_o$ &  $P_{LOS}$ constant & 20 \\
			\hline
		\end{tabular}
\end{table}

\subsubsection{Mobile UAV}
Using a mobile UAV for localizing ground users requires careful path planning to ensure a satisfactory localization accuracy and concurrently, maximize the number of users served. Path planning includes choosing flying altitude and the corresponding anchor points. Additionally, the anchor points inter-distance is an essential part of the UAV trajectory as they influence the flying time and hence the energy requirements.

To have a clear insight, in the following we investigate each design parameter individually using our simulations.
\par
\begin{figure}
\centering
     \includegraphics  [width=.5\linewidth] {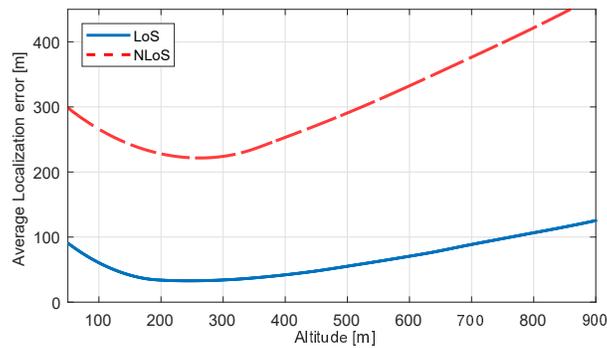} 
    \caption{Localization error in case of LoS and NLoS links}
    \label{fig:locErr}
\end{figure}

\textit{ Trajectory altitude} - Figure \ref{fig:locErr} presents the localization accuracy as a function of the UAV altitude. The figure shows that the UAV anchor outperforms the ground anchor (i.e., low altitudes) when flying at the optimal altitude. In the figure, the localization error assuming LOS and NLOS for a trajectory with $M = 3$ is illustrated. It is seen from the figure that, for both LOS and NLOS cases, the error is a convex function of $h_{ABS}$ because of the exponentially decreasing variance of the $\psi_j$ with $h_{ABS}$ \cite{7037248}. On the other hand, for large values of $h_{ABS}$, and hence $d$, the path loss curve has a decreasing slope meaning that any tiny variations in the path loss curve will lead to a large estimation error) making localization accuracy inversely related to $h_{ABS}$. Finally, the figure also shows that the localization error is always better for a LOS channel.

\par
\begin{figure}
\centering
     \includegraphics  [width=.5\linewidth] {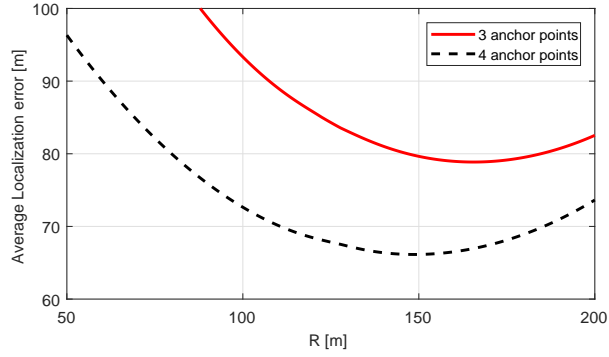} 
    \caption{Localization error for 3 and 4 anchor points at different trajectory radius.}
    \label{fig:TrajRad}
\end{figure}

 \textit{Trajectory radius ($R$)} - In Figure \ref{fig:TrajRad} we present how the trajectory radius influences the localization error performance when $h_{ABS}$ is fixed. The figure shows that high localization errors occur when $R$ is small (i.e., $R = 50$). This is due to the fact that, for multilateration, at small distances between the anchor points, a small estimation error in the distance will lead to a large error in the estimated location. Hence, increasing $R$ decreases the localization error from $150\,$m to $80\,$m in trajectory with 3 anchor points and from $95\,$m to $65\,$m in trajectory with 4 anchor points at optimal $R$. The cost here, however, is the higher energy required for larger $R$.

\par\textit{Number of anchor points ($M$)} - The localization accuracy with 3 and 4 anchor points is presented in Figure \ref{fig:TrajRad}. As shown in the figure, having more anchor points increases the localization accuracy. Nevertheless, injecting more anchor points in the trajectory implies a longer total hovering time and longer traveling distances for the UAV, leading to a higher energy consumption.

\subsubsection{Multiple Hovering UAVs}
The main disadvantage of using one mobile UAV is the delay required to reach at least 3 anchor points before estimating the position of the UE. This challenge becomes even more serious when the UE is moving. An alternate solution here is to use multiple UAVs, one at each anchor points. In this approach, at least 3 UAVs are required to cooperatively defined the location of ground users. Similar to the mobile UAV case, altitude and inter-distance of UAVs are the design parameters that control the system performance. In the case when all UAVs are flying at the same altitude, the results shown in Figure \ref{fig:locErr} will also hold for the multiple UAVs case. The trajectory radius, moreover, can be simply mapped to the UAVs inter-distance $l_M$ using
\begin{eqnarray}
l_{M} = 2 R \sin \left( \frac{\vartheta}{2} \right),
\label{trjLen}
\end{eqnarray}
where $\vartheta = \frac{2\pi}{M}$ is the angle between any two adjacent anchor points. Accordingly, there exists an optimal $l_M$ to minimize the location error with same trends presented in Figure \ref{fig:TrajRad}.
\subsection{Conclusions}
Using the detailed UAV channel models, we can conclude that systems exploiting UAV for communication or localization services have shown to benefit from the high altitude of the UAV. 
With respect to interference, a downlink UAV channel is always benefiting compared to the ground-to-ground channel. 
These results however still need to be verified using realistic simulations or experiments. 
There is a lack of semi-deterministic (i.e. using realistic environments) simulation studies beyond the statistical ones, as real UAV BS locations are not there yet, so these cannot be simulated as we did before in Section~\ref{Sec:AUE_sim}. 
With respect to experiments, it is not trivial to put BS equipment on a UAV, and obtain decent experimental evaluations here. 
\section{Research Directions and Future Work}\label{Sec:future}
In this section, we identify the directions of the future research towards more reliable UAV communications. First, we evaluate the missing channel measurements in literature and why these measurements have to be performed. Secondly, we discuss the limitations of the existing channel models and the way to improve them. Next, aerial cellular handovers and their open challenges are discussed. Afterwards we evaluate the open questions for the wireless backhaul for aerial BSs. To end, the challenges related to finite amount of energy stored on the drone are discussed, mainly, path planning in relation to the wireless connectivity and wireless power transfer.

\subsection{Channel Measurements}
In the future, UAVs will be deployed in a large variety of the environments: not only a typical urban, rural and sub-urban, but also in industrial settings (e.g. for the infrastructure inspection), over water, high-ways, stadiums and similar large public areas etc.
Moreover, indoor channel models may be needed for some specific use-cases \cite{Poza18} such as search and rescue missions, entertainment and deliveries.
\par
For the potential measurement campaigns, we believe that the listed environments must be considered.
To start, future measurement campaigns should take into account various city layouts and building dimensions such as the Manhattan grid or the more irregular layout of European cities. As a result, parametric models can be developed, using these models and the parameters of a specific city we can better model the aerial channel characteristics of that city.
\par
Secondly, there is no channel model for very crucial flight maneuvers such as take-off and landing.
Depending on the environment, the UAV can experience severe \acrshort{LS} and rich multi-path propagation conditions while executing these important tasks. Measurements during these maneuvers will give us a better understanding about the challenges these maneuvers face, when the UAV transitions between different types of channels, imposes on these systems.
\par
Another set of measurements that can be of great interest is the measurements of \acrshort{A2G} for both moving and static aerial nodes. In addition, mobile air-to-air channels measurements can be performed. This will shine a light on how mobility affect UAV communications. 
\par
Airframe shadowing and interference from on-board equipment received limited attention from the research community \cite{7986413,Bergh2015AnalysisOH}. Indeed, they have to be characterized. Typically, the emitted power is low, but it can have a large impact on the final performance due to the short distance between the interferer and the receiver. 
\par
The majority of the published works considers only the first-order statistics of the small-scale fading. Even though in \cite{6492100,8287893} the second order statistics of envelope level crossing rate and average fade duration are discussed, unfortunately, the number of works analyzing this vital issue is restricted. Another important issue that has to be investigated via measurements is the stationarity distance and its altitude dependent behavior.
\par
5G-oriented measurements such as mmWave and \acrshort{MaMIMO} are of interest as well, since it was pointed out that interference is the main issue for aerial communication using cellular networks. Both of these technologies, mmWave and MaMIMO, are key to lower the amount of interference in the aerial context. MaMIMO is able to relief the interference constraints by using dozens of antennas. These antennas make the technology capable of beam-forming the signals to specific users. On the other hand, mmWave will be able to lower the amount of interference due to its big available bandwidth that can be shared with more users. However, the behavior of these channels may be quite peculiar since even for classical \acrfull{MIMO} the fading is more severe as it was shown in \cite{Vinogradov15} and follows so-called Second Order Scattering Fading distribution \cite{Jari-06}.

\subsection{Channel Models}
The availability of high-resolution 3D maps makes possible the use of semi-deterministic channel models where some of the models parameters are extracted directly from the map (similar to the well-known COST models for cellular networks \cite{COST231}).
As far as we know, such models have not been published yet, but they do promise to be very successful to correctly model the channel characteristics of a specific environment.
\par
Potential use-cases, where UAV are used as an assistant for the autonomous cars, can require Air-to-Vehicular~(A2V) channel models. In this case, the challenging aerial channels will be paired with a high mobility, this results in a very dynamic aerial context. In order to effectively study this environment, a channel model is needed to simulate the communication between the vehicles and the UAVs.
\par
Another possible challenge is an accurate A2A models taking into account various flight situations. These flight situations range from take-off and landing to hovering and over cruising at a stable speed to making complex movements in the air.
\par
The published models are mostly dedicated to \acrshort{PL} and \acrshort{LS} modeling. However, an accurate modeling of the first- and second-order statistics of the \acrshort{SS} is needed for the future wireless network design and deployment.
Moreover, we believe that the most accurate channels model should take into account the non-stationary behavior of the \acrshort{SS}. To create such models, the quasi-stationarity distance must be estimated as it was done for indoor and vehicular channels in \cite{ Bernado-2008-URSI,Herdin-2004, 6901280,7974797, Vinogradov15}.

\subsection{Cellular Handover for Drones}
UAVs connected to ground cellular networks are normally capable of detecting several LOS BSs which are radiating their signals through their antennas main-lobe and side-lobes. Accordingly, the pattern of the cells are different in the air. This fact will result in a different handover characteristic for \acrshort{AUE}s. Such characteristics are highly dependent on the blockage distribution, height and mobility pattern of the flying UAV. As a matter of fact, more frequent handovers can be expected where some of them might fail due to a low received signal power from side-lobes. In order to establish reliable and safe cellular-connected UAVs, a qualitative and quantitative understanding of such network behavior is of utmost importance. This is still an open problem.

\subsection{Wireless Backhaul for UAV BSs}
As mentioned in section \ref{sec:UAVinfrastructure}, using UAVs as aerial communication platforms has several benefits. However, such a platform requires a wireless backhaul link, which is very challenging in comparison with the backhaul of terrestrial infrastructure. Satellite-assisted backhaul links are an expensive solution, moreover, the latency of satellite communication impedes the working of time-constrained services as real-time control and VoIP. Therefore, alternative low-latency and high-throughput solutions are needed urgently. Existing cellular network infrastructure can be reused for this purpose, which is an ongoing active research topic.

\subsection{Path Planning}
Having communication services mounted on UAVs brings extra flexibility to relevant communication systems in terms of mobility and adapting the services on-demand. However, the altitude-dependent coverage and mobility of energy-constrained UAVs introduce new challenges for the design of UAV-based wireless communications. In general, there are two constraints to consider: the ground nodes distribution and the on-board energy limitations. Consequently, the communication service is optimized according to these constraints. For instance, in \cite{Zeng} the authors provide a practical design for fixed-wing UAVs to hover around a ground UE in order to maximize the communication throughput, given its limited on-board energy. In \cite{sallouha2018energy}, the on-board energy constraints for rotary wings UAVs have been taken into account alongside the serving area in order to minimize the localization error. Nevertheless, further studies are required to jointly optimize more than one service at a time. Moreover, recently, trajectory optimization has been extended to various new promising setups that require further investigation, such as UAV-enabled data collection \cite{zhan2018energy}, multi-UAV cooperative communication \cite{wu2018joint}, \cite{liu2018comp}, and UAV-enabled wireless power transfer \cite{xu2018uav}.

\subsection{Power Allocation and Wireless Power Transfer}
The main components of energy consumption of an UAV are the communication related energy and the propulsion energy, which is required for moving and hovering. In general, the communication-related energy is ignorable in comparison to the UAV’s propulsion energy, i.e. a few watts versus hundreds of watts. In some specific applications, however, new energy requirements are introduced such as when a UAV is used for wireless power transfer and backscatter data collection. Given the different power requirements for different flying modes and the high power requirements for such applications, a careful allocation of the limited on-board power becomes even more crucial. In \cite{sallouha2018energy}, we have shown that hovering is power inefficient compared to moving with an optimal velocity. On the other hand, wireless power transfer and backscatter applications work better when hovering \cite{tekdas2009using}. Therefore, mobility management and power allocation between application requirements are becoming one of the most attractive research directions in UAV-based communications.

\section{Conclusions}\label{Sec:conc}
In this work we provided a comprehensive tutorial on the use of UAVs in wireless networks.
Contributions of this tutorial:
\begin{itemize}
\item Classification of UAV communication research topics
\item Mapping of important use cases following the UAV role in the system (aerial user equipment or aerial base station) and also according to performance requirements
\item Comprehensive overview of the channel modeling efforts suitable for UAV performance analysis
\item Analytical, simulation and experimental evaluation of UAV as UE scenario, highlighting the important impact of interference
\item Performance analysis of UAV as BS showing the high potential for this use case and motivating here more research in terms of simulations and experiments
\item Summary of the main directions to improve the channel models and performance analysis beyond the metrics already discussed in this paper.
\end{itemize}

\section{Acknowledgements}\label{Sec:ack}
\begin{description}
\item This work is part of a project that has received funding from the SESAR Joint Undertaking (JU) under grant agreement No. 763702.
The JU receives support from the European Union’s Horizon 2020 research and innovation programme and the SESAR JU members other than the Union.
\item This research was also partly funded by the Research Foundation Flanders (FWO), project no. S003817N “OmniDrone”.
\end{description}
\bibliography{main.bib}
\bibliographystyle{IEEEtran}

\begin{IEEEbiographynophoto}{Evgenii Vinogradov}
received the Dipl. Engineer degree in Radio Engineering and Telecommunications from Saint-Petersburg Electro-technical University (Russia), in 2009. After several years of working in the field of mobile communications, he joined UCL (Belgium) in 2013, where he obtained his Ph.D. degree in 2017. His doctoral research interests focused on multidimensional radio propagation channel modeling. In 2017, Evgenii joined the electrical engineering department at KU Leuven (Belgium) where he is working on wireless communications with UAVs and UAV detection.
\end{IEEEbiographynophoto}

\begin{IEEEbiographynophoto}{Hazem Sallouha}
received the B.Sc. degree in electrical engineering from Islamic University of Gaza (IUG), Palestine, in 2011, the M.Sc. degree in electrical engineering majoring in wireless communications from Jordan University of Science and Technology (JUST), Jordan in 2013. Currently he is a Ph.D. candidate at the Department of Electrical Engineering, KU Leuven, Belgium. His main research interests are localization techniques, communications with unmanned aerial vehicles, Internet of things networks and machine learning algorithms for localization.
\end{IEEEbiographynophoto}
\begin{IEEEbiographynophoto}{Sibren De~Bast}
obtained his B.Sc. degree in Electrical Engineering and Computer Science from KU Leuven, Belgium in 2015. Afterwards, he gained his M.Sc. degree in Electrical Engineering from KU Leuven in 2017. During his master thesis, he researched wireless streaming of stereoscopic video from aerial nodes to a ground station, with the focus on low latency. Currently, he is pursuing the Ph.D. degree at KU Leuven. He is focusing on AI-enhanced wireless communication with UAVs, regarding Massive MIMO and low-latency, high-throughput connections.
\end{IEEEbiographynophoto}
\begin{IEEEbiographynophoto}{Mohammad Mahdi Azari}
received his B.Sc. and M.Sc. degrees in electrical engineering from University of Tehran, Tehran, Iran. He is currently a doctoral research fellow at the department of electrical engineering, KU Leuven. He worked for Huawei prior to KU Leuven. His current research interests include hybrid terrestrial-aerial communication networks, 5G and heterogeneous cellular networks, cooperative communications, and coding and network coding theory.
\end{IEEEbiographynophoto}
\begin{IEEEbiographynophoto}{Sofie Pollin}
obtained her PhD at KU Leuven in 2006. She continued her research on wireless communication at UC Berkeley. In November 2008 she returned to imec to become a principal scientist in the green radio team. Since 2012, she is professor at the electrical engineering department at KU Leuven. Her research centers around Networked Systems that require networks that are ever more dense, heterogeneous, battery powered and spectrum constrained. She has been working on drone communication since 2012, given various invited talks on the topic, and authored invited book chapters, journals and tutorials related to UAV communication. She is also co-founder of the ACM workshop DroNET, focusing on drone communication and networks. Prof. Pollin has experience with tutorials at academic conferences such as ICC or Crowncom, or mixed industry/academic fora such as Embedded Silicon West or the SDR forum.
\end{IEEEbiographynophoto}





\end{document}